\documentclass[a4paper,11pt]{article}
\usepackage[position=t,singlelinecheck=off]{subfig}
\usepackage{amssymb,amsmath}
\usepackage{graphics,graphicx,float}
  \usepackage{mathabx}
 \usepackage{epstopdf}

\def\beq{\begin{equation}}
\def\eeq{\end{equation}}
\def\bea{\begin{eqnarray}}
\def\eea{\end{eqnarray}}
\def\nn{\nonumber}

\def\lo{\left(}
\def\rc{\right)}

\makeatletter
 
  \def\@cite#1#2{${\mbox{#1\if@tempswa , #2\fi}}$}
\makeatother

\begin{document}
\thispagestyle{empty}
\vspace*{3cm}
\begin{center}
{\LARGE\sf Quasi-Bell states in a strongly coupled qubit-oscillator system and their delocalization in the phase space}\\

\bigskip\bigskip
R. Chakrabarti and B. Virgin Jenisha

\bigskip
\textit{
 Department of Theoretical Physics, 
University of Madras, \\
Maraimalai Campus, Guindy, 
Chennai 600 025, India \\}
\end{center}

\vfill
\begin{abstract}
 We study the evolution of bipartite entangled quasi-Bell states in a strongly  coupled 
 qubit-oscillator system in the presence of a static bias, and extend it to  the ultra-strong coupling regime. Using the adiabatic approximation the reduced density matrix of the qubit is obtained for the strong coupling domain in \textit{closed form} that 
 involves linear combinations of the Jacobi theta functions. 
 The reduced density matrix of the oscillator yields the phase space Husimi $Q$-distribution. In the strong coupling regime the $Q$-function evolves to uniformly separated macroscopically distinct Gaussian peaks representing `kitten' states at certain specified times that depend  on  \textit{multiple} time scales present in the interacting system. For the ultra-strong coupling realm  the delocalization in the phase space of the oscillator is studied by using the Wehrl entropy and the complexity of the quantum state. For a small phase space amplitude the entangled quasi-Bell state develops, during its time evolution, squeezing property and nonclassicality of the photon statistics which are  measured by the quadrature variance and the Mandel parameter, respectively.
\end{abstract} 
 
\newpage
\setcounter{page}{1}

\section{Introduction}
One of the ubiquitous and most important models in the quantum physics describes a  two-level system (qubit) interacting with a simple harmonic oscillator. In quantum optics it describes the interaction of an atom with an electromagnetic field mode in a cavity. For a small detuning between the oscillator and the qubit frequencies, and also for a weak  qubit-oscillator coupling, the dynamical behavior of the interacting system is accurately described by the exactly solvable Jaynes-Cummings model [\cite{JC1963}] that employs the rotating wave approximation.  Recently, however, a variety of experimental situations pertaining to the stronger coupling domain, 
where the rotating wave approximation no longer holds, have been investigated.
Various experimental realizations such as a nanomechanical resonator capacitively 
coupled to a Cooper-pair box [\cite{ABS2002}], a quantum semiconductor microcavity
undergoing excitonic transitions [\cite{ALT2009}], a flux-biased superconducting quantum  circuit that uses large nonlinear inductance of the Josephson junctions to achieve
ultrastrong coupling with a coplanar waveguide resonator  [\cite{Niemczyk2010}] have been  achieved.
The superconducting circuits based on the Josephson junctions behave, in certain parametric range, as effective two level systems, and can exhibit macroscopic quantum coherence [\cite{YN2005}, \cite{YN2011}]. These artificial atoms  have advantage  in that  they  offer additional flexibility in selecting the control parameters which  facilitate experimental tests of fundamental quantum mechanical principles at a macroscopic scale [\cite{NJBN2012}]. The tailor-made physical characteristics of the superconducting circuits and their easy adjustability, say by changing  external magnetic fields, make them particularly suitable candidates for the quantum simulators 
[\cite{BAN2011}-\cite{GAN2014}]. Integrated hybrid quantum circuits involving atoms, spins, cavity photons, and superconducting qubits with nanomechanical resonators are also of much interest [\cite{ZAYN2013}].

\par

Moving beyond the rotating wave approximation in these strongly interacting systems one needs to incorporate terms in the qubit-oscillator interaction Hamiltonian that do not preserve the total excitation number. In the parametric regime where the oscillator frequency dominates the qubit frequency the authors of Refs. [\cite{IGMS2005},  \cite{AN2010}]  have
advanced an adiabatic approximation scheme that utilizes  the disparity of the  time scales of the qubit and the oscillator. The dynamical state of the fast-moving  oscillator is assumed to instantaneously adjust to the slow-changing state of the qubit. This facilitates decoupling of the full Hamiltonian into sectors related to each time scale, and allows approximate evaluation [\cite{IGMS2005}] of  eigenstates of the system. In addition, an analytical expression of the 
time-dependent behavior of the two level system has been obtained [\cite{IGMS2005}] in a relatively weak coupling limit. Studying a system of two qubits coupled to a single oscillator mode in the context of the Tavis-Cummings model the authors of  
Ref. [\cite{ARE2012}] have analytically evaluated the collapse and revival dynamics of the qubit reduced density matrix elements. Their procedure employs the Poisson sum formula [\cite{AAR1999}] that leads to expressions of the said elements as infinite sums. Following the extensive experimental developments [\cite{YN2005}, \cite{YN2011}] for the superconducting  qubits, a static bias term has been included  in the qubit Hamiltonian and the corresponding energy spectrum has been investigated [\cite{AN2010}, \cite{HG2010}]. A time-dependent bias acting as a control parameter 
has also been considered [\cite{SAN2010}].

\par

For the coupled qubit-oscillator system the entangled quasi-Bell states are found to be of much interest. Usually known as the Schr\"{o}dinger cat states they exhibit entanglement of the microscopic atomic states and the photonic coherent states  that can be regarded as macroscopic for reasonably large values of the coherent state amplitudes. 
  In the context of quantum information theory the hybrid entanglement 
involving a discrete and a continuous quantum variable may offer advantages such as achieving 
near-deterministic quantum teleportation [\cite{LJ2013}]. Moreover, hybrid entanglement lies at the heart of the elegant quantum bus [\cite{Milburn2008}] approach where direct qubit-qubit interactions are avoided since a common continuous mode plays the mediating role. The Bell inequality tests involving these qubit-field entangled states have been proposed [\cite{PSSAJ2012}] recently. Realization of the micro-macro entangled states via a controllable interaction  of a single-mode microwave
cavity field with a superconducting charge qubit, and the subsequent
creation of  the superposition of  macroscopically distinguishable field modes by virtue of  measuring  the charge states of the qubit have been proposed [\cite{LWN2005}].  Much experimental activity [\cite{A2014}] is currently aimed at demonstrating  such hybrid entanglement. Generation  
 of  the optical hybrid entanglement by the quantum superposition of non-Gaussian operations 
on distinct modes has been achieved [\cite{Jeong2014}]. Also following a measurement based approach, the hybrid entanglement between two remote nodes living in the Hilbert spaces of different dimensionality has been established [\cite{Morin2014}]. The superconducting circuits have also been used for controllable and deterministic generation of complex superposition of states [\cite{LWN2004}-\cite{Hofheinz2009}].

\par

Our objective in the present work is twofold. Within the adiabatic approximation scheme we analytically study the time dependence of quasi-Bell states in a coupled qubit-oscillator system endowed with a dominant oscillator frequency and an external static bias of the qubit. Our study includes both the strong and the ultra-strong coupling domains. 
Starting with a quasi-Bell bipartite initial state we obtain the  evolution of the reduced density matrices of the qubit and the oscillator, respectively. The qubit  density matrix   provides the von Neumann entropy of the system that measures the entanglement and the  mixedness of the state. Proceeding further, the oscillator reduced density matrix that yields the Husimi $Q$-function [\cite{Schleich2001}] on the phase space is studied. In the strong coupling regime we retain up to the quadratic terms in the Laguerre polynomials in the phase factors and analytically express the reduced density matrix elements in closed forms  via the Jacobi theta functions.   
To evaluate the analytical expressions we treat the bias perturbatively, and retain terms up to second order in the expansion parameter. The summation procedure may, however, be extended to an arbitrary order. The $Q$-distribution provides the variance of the quadrature variables of the oscillator as linear combinations of the Jacobi theta functions.  We will later specify the parametric range where our theta function based construction is valid. In its region of validity the leading terms in the theta function structure provides simple estimates, say, for the revival time for the off-diagonal qubit matrix element. Analysing the $Q$-function in  the Kerr-like nonlinear self-interacting photonic models
it has been observed [\cite{MTK1990}-\cite{MBWI2001}] that an initial coherent state therein evolves, at certain specified times, to macroscopic superposition of coherent states popularly known as `kitten' states. In the current 
qubit-oscillator interacting model the reduced density matrix of the oscillator shows similar properties  with, however, an important distinction that stems from the existence  of 
interaction-generated  multiple time scales.      

\par

In the ultra-strong coupling regime we employ the  long range time  dependence of the  Wehrl entropy [\cite{W1978}] and the complexity [\cite{SA2002}] of the  
$Q$-distribution  that  measure the delocalization  of the oscillator  on the phase space.  An analytical evaluation of the complexity of the  coupled qubit-oscillator system is developed. Both of these measures of  delocalization  are found to be qualitatively similar. 
The  time dependence of the  Wehrl entropy in the ultra-strong coupling domain exhibits, after a transient period of rapid build up, a quasi-stationary behavior modulo the random high frequency fluctuations. We estimate the initial production-time of the Wehrl entropy. For small values of the  amplitude of the quasi-Bell state on the phase space, interference of modes produces the squeezing effect. Using the Mandel parameter [\cite{M1979}] the nonclassicality of the photon statistics is also observed in this instance.

\section{The reduced density matrices}
\label{General}
\setcounter{equation}{0}
We study a coupled qubit-oscillator system with the  Hamiltonian [\cite{IGMS2005},  \cite{AN2010}] that reads in natural units $(\hbar=1)$ as follows:
\beq 
 H = -\frac{\Delta}{2} \sigma_x - \frac{\epsilon}{2} \sigma_z +
 \omega a^{\dagger}\, a + \lambda  \sigma_z \,(a^{\dagger} + a),
 \label{Hamiltonian}
\eeq
where the harmonic oscillator with a frequency $\omega$ is described by the raising and lowering operators $(a^{\dagger}, \,a| \hat{n} \equiv a^{\dagger} a)$, and the qubit characterized by an energy splitting $\Delta$ as well as an external static bias $\epsilon $ is expressed via the spin variables $(\sigma_x, \,\sigma_z)$. The qubit-oscillator coupling strength is denoted by $\lambda$.  The Fock states 
$\{\hat{n} |n\rangle = n |n\rangle,\,n = 0, 1,\ldots;\;a \,|n\rangle = \sqrt{n}\,|n - 1\rangle, 
a^{\dagger}\, |n\rangle = \sqrt{n + 1}\,|n + 1\rangle\}$ provide the basis for the oscillator, and the eigenstates  $\sigma_z |\pm 1\rangle = \pm \,|\pm 1\rangle$ span the space of the qubit.  The Hamiltonian (\ref{Hamiltonian}) is not known to be exactly solvable.
In the present work we follow the adiabatic approximation [\cite{IGMS2005},  \cite{AN2010}] that hinges on the separation of the time scales governed by the high oscillator frequency and the (renormalized) low qubit frequency: 
$\omega \gg \Delta$.  

\par

To facilitate our construction of the evolution of the quasi-Bell states, we, following [\cite{IGMS2005}], now give a short review
of the diagonalization process of the Hamiltonian (\ref{Hamiltonian}) in the adiabatic approximation scheme. The high-frequency oscillator is assumed to instantaneously adjust to
the slow-changing  state of the qubit observable $ \sigma_z$ so that the construction permits, in the course of   diagonalization of the oscillator mode, replacing the spin-variable 
 $\sigma_z$ with its eigenvalue: $\pm 1$. The effective Hamiltonian of the harmonic oscillator now reads
\beq
 H_{\cal O} = \omega a^{\dagger} a \pm \lambda   (a^{\dagger} + a).
\label{H_photon}
\eeq
The number states which are shifted symmetrically by the displacement operator diagonalize the above Hamiltonian: 
\beq
 |n_{\pm} \rangle=\mathrm{D}^{\dagger}\left(\pm \lambda/\omega \right) | n \rangle,\qquad
 \mathrm{D}(\alpha)= \exp(\alpha a^{\dagger}-\alpha^*a),
 \label{D_state}
\eeq
where the degenerate eigenenergies are as follows:
\beq
 E_n^{\pm} \equiv  {\cal E}_{n} = n\omega-\frac{\lambda^2}{\omega}.
 \label{E_shifted}
\eeq
Identically displaced states  maintain orthonormality: $\langle m_{\pm}| n_{\pm}\rangle 
= \delta_{m, n}$, whereas the overlap between the Fock states displaced in opposite directions is given by [\cite{IGMS2005}] 
\beq
  \langle m_-| n_+ \rangle =
 \begin{cases}
 (-1)^{m-n}\;\exp \lo-\frac{x}{2} \rc  x^{\frac{m-n}{2}} \sqrt{n!/m!} \, \, L_n^{m-n}\lo x\rc 
 & \forall \,m \geq n,\\
 \exp \lo-\frac{x}{2} \rc x^{\frac{n-m}{2}} \sqrt{m!/n!} \, \, L_m^{n-m}\lo x\rc & \forall \,m < n.
\end{cases}
\label{overlapp}
\eeq
In (\ref{overlapp}) the associated Laguerre polynomial reads $L_{n}^{j}(x) = \sum_{k = 0}^{n} 
\,(-1)^{k}\, \binom{n + j}{n - k}\,\frac{x^{k}}{k!}$, and the parameter $x = \left( 2\lambda/ \omega\right)^2$ acts as  the perturbative expansion parameter.

\par

After diagonalizing the high frequency oscillator component of the Hamiltonian, one now turns attention to the low frequency qubit part. Tensoring the qubit states with the displaced oscillator basis states
$\{|\pm 1,n_{\pm}\rangle\,|\, \forall n = 0, 1, \ldots\}$ the corresponding matrix elements of the Hamiltonian may now be readily constructed.  For a dominant oscillator frequency $\Delta/\omega \ll 1$ one may neglect [\cite{IGMS2005}] the matrix elements that mix the oscillator states with different eigenvalues of its number operator. This reduces the Hamiltonian to a block-diagonal form where each block mixes the displaced oscillator states with identical number of photons.  
The Hamiltonian for the $n$-th block assumes the form:
\beq
H=
\begin{pmatrix}
n \omega - \dfrac{\lambda^2}{\omega} - \tilde{\epsilon} & \delta_n \\
\delta_n & n \omega - \dfrac{\lambda^2}{\omega} + \tilde{\epsilon}
\end{pmatrix},
\;\;\delta_n = -\dfrac{\widetilde{\Delta}}{2}  L_{n}(x), \;\widetilde{\Delta} = \Delta \exp \lo - \frac{x}{2}\rc,  \; \tilde{\epsilon}=\dfrac{\epsilon}{2}
\label{H-block}
\eeq
that may be diagonalized in the basis of the  eigenstates listed below
\beq
|{\cal E}_n^{\pm} \rangle =  \sqrt{\dfrac{\chi_n \mp \tilde{\epsilon}}{2\chi_n}} | 1,n_+ \rangle \pm \dfrac{\delta_n}{|\delta_n|} \sqrt{\dfrac{\chi_n \pm \tilde{\epsilon}}{2\chi_n}}| -1,n_- \rangle,
\label{Hn_eigenstate}
\eeq
where the respective eigenvalues read
\beq
{\cal E}_{n}^{\pm} = n \omega - \lambda^2/\omega \pm \chi_n,\qquad
\chi_n=\sqrt{\delta_n^2+\tilde{\epsilon}^2}.
\label{Hn_eigenvalue}
\eeq

\par

Given the above construction of the eigenstates of the system we now study the time evolution of the qubit-oscillator entanglement in the presence of a bias ($\epsilon \neq 0$). 
The quasi-Bell initial states of the coupled system read
\beq
|\psi(0)\rangle^{(\pm)} = \dfrac{1}{\sqrt{2}}\left(|1,\alpha \rangle \pm |-1, -\alpha \rangle\right),\qquad
|\alpha \rangle = \mathrm{D}(\alpha) |0 \rangle,
\label{quasi_Bell}
\eeq
where the coherent state  $\{|\alpha \rangle \; \forall \alpha = \mathrm{Re}(\alpha) + i\, \mathrm{Im}(\alpha) \in \mathbb{C}\}$ for the oscillator degree of freedom is realized by the action of the 
displacement operator on the vacuum.
The  evolution of the initial state (\ref{quasi_Bell}) is given by
\bea
 |\psi(t)\rangle^{(\pm)} &=& \dfrac{1}{\sqrt{2}}  \exp\left(-\dfrac{1}{2}|\alpha_{+}| ^2 -i\left(\mathrm{Im}{(\alpha)}\dfrac{\lambda}{\omega}+{\cal E}_n t\right )\right) \;\times \nn \\
  &&\times \sum_{n=0}^{\infty} 
\dfrac{1}{\sqrt{n !}} \alpha_{+}^n
 \left[ \Xi _n^{\pm} \exp(-i\chi_nt)
 | {\cal E}^{+}_n  \rangle \mp (-1)^n \dfrac{\delta_n}{|\delta_n|} \, \, \Xi _n^{\mp} \exp(i \chi_n t) | {\cal E} _n^{-} \rangle \right], \qquad
\label{state_t} 
\eea
where $\alpha_{+}=\alpha + \lambda /\omega$,  and $ \Xi_n^{\pm}=\sqrt{\dfrac{\chi_n -
 \tilde{\epsilon}}{2\chi_n}} \pm (-1)^n \dfrac{\delta_n}{|\delta_n|} \sqrt{\dfrac{\chi_n + \tilde{\epsilon}}{2\chi_n}}$.
 On the rhs of (\ref{state_t}) and hereafter the upper and the lower signs refer to respective quasi-Bell states given in (\ref{quasi_Bell}).
 The construction of  the state (\ref{state_t}) readily yields the evolution of the density matrix of the composite bipartite system: 
\beq
\rho^{(\pm)}(t) = |\psi(t)\rangle^{(\pm)}\langle \psi(t)|.
\label{density_composite}
\eeq
The marginalized density matrices for the individual degrees of freedom may be obtained by
partial tracing  of the composite density matrix (\ref{density_composite}). For instance, tracing of the oscillator degree of freedom yields the reduced density matrix of the qubit. 
Its explicit structure obtained via (\ref{density_composite}, \ref{state_t}) assumes the form 
\beq
\rho_{\cal Q}^{(\pm)}(t) \equiv \hbox {Tr}_{\cal O}\,\rho^{(\pm)}(t)=
\begin{pmatrix}
\dfrac{1}{2} \mp \zeta & \pm \xi^{(\pm)} \\
\pm (\xi^{(\pm)})^* & \dfrac{1}{2} \pm \zeta
\end{pmatrix},
\label{density_matrix}
\eeq
where the components read
\bea
\zeta &=& \tilde{\epsilon}\; \exp \left(-|\alpha_{+}|^2\right) \sum_{n=0}^{\infty} (-1)^n\, \dfrac{|\alpha_{+}|^{2n}}{n!}\,\delta_n\,  \dfrac{\sin ^2 \chi_n t}{\chi_n^2},
\label{density_diagonal} \\
\xi^{(\pm)} &=&  \dfrac{1}{2}\; \exp(-|\alpha_{+}|^2) \sum_{n, m = 0}^{\infty} (-1)^m\, \dfrac{({\alpha_{+}})^{n}\,({\alpha_{+}}^{*})^{m}}{\sqrt{n!\, m!}}\, 
{\mathsf{C}_n^{\pm}}(t)\, \mathsf{C}_m^{\mp}(t)\quad\times\nn \\
& &\times \;\; \exp \big (-i (n - m)\omega t \big)\; \langle m_- | n_+\rangle, 
\label{density_offdiagonal}
\eea
and the coefficients are given by
\beq
\mathsf{C}_n^{\pm}(t)=  \mathsf{A}_n^{\mp} \exp (i \chi_n t)+\mathsf{B}_n^{\pm} \exp (-i \chi_n t),
\label{o_density_c}
\eeq
\beq
 \mathsf{A}_n^{\pm}= \dfrac{\chi_n+\tilde{\epsilon} \pm (-1)^n \delta_n}{2 \chi_n}, \quad 
\mathsf{B}_n^{\pm}= \dfrac{\chi_n-\tilde{\epsilon} \pm (-1)^n \delta_n}{2 \chi_n}.
\label{o_density_ab}
\eeq
It follows from (\ref{density_diagonal}) that in the bias-free limit $(\epsilon = 0)$ the diagonal elements
of the reduced density matrix  (\ref{density_matrix}) of the  qubit are time-independent. In the said limit it follows 
\beq
\mathsf{C}_n^{\pm}(t)|_{\epsilon = 0} = \cos \delta_{n} t \mp i (-1)^{n} \sin \delta_{n}t = \exp \big(\mp i (-1)^{n} \delta_{n} t\big).
\label{C_biasfree}
\eeq

\par
 
In general, the pair of eigenvalues of the qubit density matrix (\ref{density_matrix})  
\beq
\frac{1}{2} + \varpi^{(\pm)},\;\frac{1}{2} - \varpi^{(\pm)},\qquad \varpi^{(\pm)} = \sqrt{\zeta^2+|\xi^{(\pm)}|^2}
\label{rho_eigenvalue}
\eeq
allows us to compute its von Neumann entropy  $S(\rho_{\cal Q}) \equiv - \hbox {Tr} (\rho_{\cal Q} \, \log \rho_{\cal Q})$ as
\beq
S^{(\pm)}_{\cal Q} = - \left(\frac{1}{2} + \varpi^{(\pm)}\right) \log \left(\frac{1}{2} + \varpi^{(\pm)}\right) - \left(\frac{1}{2} - \varpi^{(\pm)}\right) \log \left(\frac{1}{2} - \varpi^{(\pm)}\right).
\label{entropy_defn}
\eeq

\par
  
Similarly the trace over the qubit basis produces the reduced density matrix of the oscillator: 
\beq
\rho_{\cal O}^{(\pm)}(t) \equiv \hbox {Tr}_{\cal Q}\,\rho^{(\pm)}(t)
\label{o_density}
\eeq
that may be expressed, via (\ref{state_t}), in terms of the displaced number states as follows:
\bea
\rho_{\cal O}^{(\pm)}(t)&=&\dfrac{1}{2} \exp(-|\alpha_{+}|^2)\sum_{n,m=0}^{\infty} 
  \dfrac{(\alpha_{+})^{n} \,  (\alpha_{+}^*)^{m}}{\sqrt{n!m!}} 
 \Big ({\mathsf{C}_n^{\pm}(t)\,\mathsf{C}_m^{\pm}}(t)^* |n_+ \rangle \langle m_+ |\nn \\
&& + (-1)^{n+m}
{\mathsf{C}_n^{\mp}}(t)^* \,\mathsf{C}_m^{\mp}(t) |n_- \rangle \langle m_- | \Big) \exp \big (-i (n - m)\omega t \big).
\label{o_density_matrix}
\eea
Apart from employing the adiabatic approximation \textit{no other} simplification has been made for  deriving  the reduced density matrix elements
(\ref{density_diagonal}, \ref{density_offdiagonal}, \ref{o_density_matrix}), which will later serve as benchmarks for checking the accuracy of the corresponding closed form evaluations.

\par

 To obtain measures of the entanglement and the mixedness of the  bipartite system  the von Neumann entropy of the reduced density matrix of the qubit  may be considered.  It is well-known [\cite{AL1970}] that
if  a composite system, comprising of two subsystems, resides in a pure state, the entropies of both subsystems are equal. In the present example it holds for the oscillator with an infinite dimensional Hilbert space and the  two-level qubit interacting with it: $S^{(\pm)}_{\cal Q} = S^{(\pm)}_{\cal O} \equiv S^{(\pm)}$.

\section{The qubit density matrix}
\label{qubit}
\setcounter{equation}{0}
We now  focus towards obtaining a closed form analytical evaluation of the qubit density matrix in the parametric regime where the oscillator frequency sets the dominant scale and the following hierarchy is maintained: $\epsilon < \Delta < \omega$. 
In the perturbative expansion given below we retain up to the second order terms 
 in the bias parameter: $O(\epsilon^{2})$. The higher order terms in  
$\epsilon$ may, however, be systematically  taken into account in our perturbative scheme. Furthermore, in the following  evaluation, we investigate  the first few collapses and revivals  of the density matrix elements.  Consequently, for the purpose of simplicity, we neglect $\epsilon$ dependent terms in phases as they will be important only in the long time limit. 
  Previous attempts in analytical evaluation of the elements of the density matrices include Refs. [\cite{IGMS2005}, \cite{ARE2012}].  Reflecting the weak coupling limit $\lambda \ll \omega$, the summation technique adopted in [\cite{IGMS2005}] retains terms $O(x) \sim O((\lambda/\omega)^2)$.  For analytically evaluating the density matrix elements in the context of the Tavis-Cummings model at higher values of the qubit-oscillator coupling strength
 the authors of Ref. [\cite{ARE2012}] have  
developed a method based on the Poisson sum formula. They have retained contributions  
  up to the quadratic terms in the expansion of the Laguerre polynomials: $L_{n}(x) = 1 - n x + \frac{1}{2!}\, \binom {n}{2}\,x^2 + O(x^3)$. 

\par

\begin{figure}
 \subfloat[]{ \includegraphics[width=8cm,height=3.4cm]{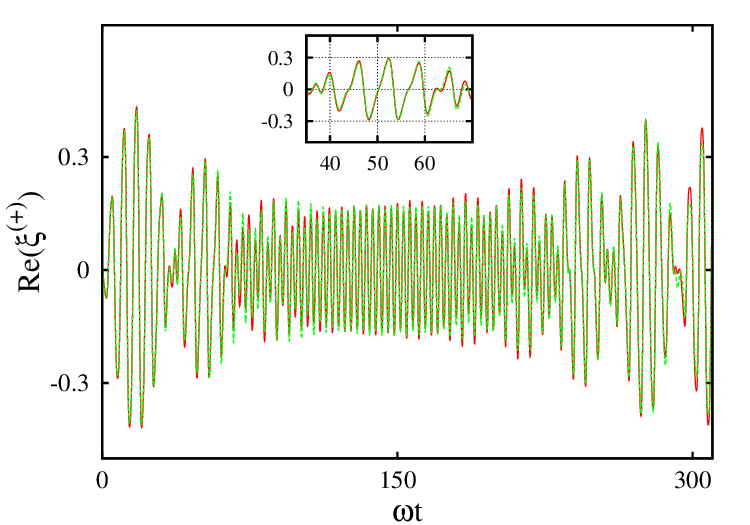}} \quad
   \captionsetup[subfigure]{labelformat=empty}
\subfloat[(b)]{\includegraphics[width=8cm,height=3.4cm]{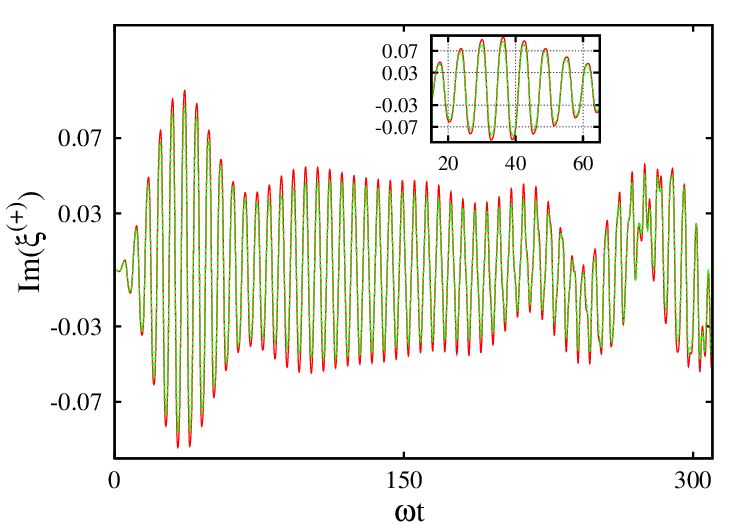}}
\caption{The evaluation of the (a) real and (b) imaginary parts of the off-diagonal component $ \xi^{(+)} $ of the density matrix for the values  $\lambda=0.15 \,\omega, \Delta=0.15 \,
\omega ,\epsilon=0.01 \, \omega $ and $ \alpha=2 $  using the series (\ref{density_offdiagonal}) (red solid) that provides exact result \textit{within the adiabatic  scheme}, and  the theta function (\ref{xi_theta}) (green dashed) based result. In this instance and hereafter we study evolution w.r.t. the \textit{scaled time} $\omega t$.}
\label{Off_diagonal_compt}
\end{figure}

We now provide an alternate
method of summing the series in  (\ref{density_diagonal}, \ref{density_offdiagonal}) by employing the Jacobi theta functions. This method has an additional benefit in that it expresses the matrix elements in a closed form. The series expansions of the Jacobi theta functions [\cite{AAR1999}]  relevant in our case  depend on two complex arguments $(q,z)$:     
\beq
\vartheta_3(q,z) = \sum_{n=-\infty}^{\infty} q^{n^2} \exp(2inz), \qquad \vartheta_4(q,z) = \sum_{n=-\infty}^{\infty} (-1)^n q^{n^2} \exp(2inz).
\label{theta_3_4}
\eeq
The series sums in (\ref{theta_3_4}) are convergent for the choice $\mathrm{Re}(q) < 1$. In our
summation scheme we retain contributions $O(x^2)$ in the density matrix elements and consider up to quadratic terms in the Laguerre polynomials.
To proceed, we, following [\cite{ARE2012}], approximate at a suitably large value of 
$|\alpha_{+}|^{2}$ the  Poisson distribution in (\ref{density_diagonal}, \ref{density_offdiagonal})
with the corresponding Gaussian limit:
\beq
P(n) \equiv \exp \left(-|\alpha_{+}|^2\right)\,\dfrac{|\alpha_{+}|^{2n}}{n!}
\rightarrow \dfrac{1}{\sqrt{2 \pi |\alpha_{+}|^2}} \exp \left( - \dfrac{\left( n-|\alpha_{+}|^2 \right)^2}{2 |\alpha_{+}|^2}  \right).
\label{gaussian_limit}
\eeq  
For a large value of $|\alpha_{+}|^2$
the lower limit of the series in $(\ref{density_diagonal}, \ref{density_offdiagonal})$ may be extended to 
$(-\infty)$. Due to the sharp decay of the Gaussian function for the parametric regime considered here the error encountered is quite small. This process is similar to the evaluation [\cite{ARE2012}] of Fourier components via the Gaussian integration technique in the Poisson-sum method, where the lower limit of integration is extended to $(-\infty)$.  We note that a truncated series at the lower limit in the sum (\ref{theta_3_4}) \textit{remains convergent} in the domain $\mathrm{Re}(q) < 1$, and it may be viewed as a close analog of the theta function. Since the difference between these  alternate cases in our parametric regime is quite small we retain the standard definition (\ref{theta_3_4}).  The resultant closed form expressions for the elements of the density matrix (\ref{density_matrix}) are now given by  
\beq
\zeta =- \dfrac{\epsilon}{2 \widetilde{\Delta}} 
\Big[\exp\Big(-2|\alpha_{+}|^2\Big)\Big( 1-f(1+x)+\dfrac{3}{4}f^2\Big) -  \dfrac{\exp\Big(-|\alpha_{+}|^2/2\Big) }{\sqrt{2 \pi |\alpha_{+}|^2}} \mathrm{Re}\big \lgroup \Theta(\tau) \big \rgroup\Big], 
\label{zeta_theta}
\eeq
\bea
\xi^{(\pm)} &=&  \dfrac{1}{2} \left(1+f+\dfrac{f^2}{4}\right)\exp\left(-2|\alpha_{+}|^2-\dfrac{x}{2}\right) + \dfrac{\epsilon}{\Delta}\,\zeta + \dfrac{\exp(-(x+|\alpha_{+}|^2)/2)}{ \sqrt{2 \pi |\alpha_{+}|^2}} 
\Big (x^{\frac12}  \,
 \mathcal{R}_{1} \times \nn\\
& &  \times \; \mathrm{Re}\Big \lgroup \varphi_{\raisebox{-1pt}{\tiny 1,0}}  \vartheta_4(q,z_1)+
 \dfrac{f}{2} \varphi_{\raisebox{-1pt}{\tiny 3,1}}\vartheta_4(q,z_3) \Big \rgroup + \frac{x}{2} \mathcal{R}_{2} \mathrm{Re}\Big \lgroup \Big(\widetilde{\varphi}_{\raisebox{-1pt}{\tiny 1}} +  \dfrac{ f }{3}\widetilde{\varphi}_{\raisebox{-1pt}{\tiny 3}}  \Big)
 \vartheta_4(\mathfrak{q},\mathfrak{z}_{2}) \Big \rgroup \nn\\ 
& &   
+ \frac{x^{\frac32}}{6} \mathcal{R}_3 \; \mathrm{Re}\lgroup \varphi_{\raisebox{-1pt}{\tiny 3,3}}
 \vartheta_4(q,z_3) \rgroup + \dfrac{x^2}{24} \mathcal{R}_4 \;
\mathrm{Re}\lgroup  \widetilde{\varphi}_{\raisebox{-1pt}{\tiny 3}}^{\raisebox{1pt}{\tiny 2}} \vartheta_4(\mathfrak{q},\mathfrak{z}_{4}) \rgroup -i \dfrac{\epsilon}{2\widetilde{\Delta}}  
 \mathrm{Im}\big \lgroup \varphi_{\raisebox{-1pt}{\tiny 0,0}}  \vartheta_4(q, z_0) \big \rgroup  \nn \\
&&  +\dfrac{ \epsilon^2}{\widetilde{\Delta}^2} \Big[ x  \mathcal{R}_2 \;
\mathrm{Re}\Big \lgroup \Big (\frac12+x \Big) \varphi_{\raisebox{-1pt}{\tiny 2,2}} \vartheta_4(q,z_2)  - \frac{5 f}{6}
 \varphi_{\raisebox{-1pt}{\tiny 4,3 }}\vartheta_4(q,z_4)   -    \Big( \Big(\frac{1}{2}+x\Big ) \widetilde{\varphi}_{\raisebox{-1pt}{\tiny 1}} \nn \\
&&
 - \frac{5 f }{6}  \widetilde{\varphi}_{\raisebox{-1pt}{\tiny 3}} \Big) \vartheta_4(\mathfrak{q},\mathfrak{z}_{2})\Big \rgroup
+ \dfrac{x^2}{24}
 \mathcal{R}_4  \;\mathrm{Re}\lgroup   \varphi_{\raisebox{-1pt}{\tiny 4,6 }}  \vartheta_4(q,z_4) 
  - \widetilde{\varphi}_{\raisebox{-1pt}{\tiny 3}}^{\raisebox{1pt}{\tiny 2}} \vartheta_4(\mathfrak{q},\mathfrak{z}_{4}) \rgroup \Big] \pm i \dfrac{\epsilon}{\widetilde{\Delta}} 
\Big [  x^{\frac12} \, \mathcal{I}_1 \times \nn \\
&& \times
\mathrm{Re}\Big \lgroup \Big ( 1+ \frac{x}{2} \Big ) \varphi_{\raisebox{-1pt}{\tiny 1,0}} \vartheta_3(q,z_1)+
 \dfrac{f}{2}  \varphi_{\raisebox{-1pt}{\tiny 3,1 }}\vartheta_3(q,z_3)  -\Big( \Big(1+\frac{x}{2} \Big) 
\widetilde{\varphi}_{\raisebox{-1pt}{\tiny 0}}^{\frac12} +\frac{f }{2} 
\widetilde{\varphi}_{\raisebox{-1pt}{\tiny 2}}^{\frac12}  \Big ) \times \nn \\
&&   \times
  \vartheta_3(\mathfrak{q},\mathfrak{z}_{1}) 
 \Big \rgroup + \dfrac{x^2}{2} \mathcal{I}_2 \;   
\mathrm{Re} \lgroup   \varphi_{\raisebox{-1pt}{\tiny 2,2  }} \vartheta_3(q,z_2)
- \widetilde{\varphi}_{\raisebox{-1pt}{\tiny 1}} \vartheta_3(\mathfrak{q},\mathfrak{z}_{2})
  \rgroup + \dfrac{x^{\frac{3}{2}}}{6} \mathcal{I}_3   \mathrm{Re}\lgroup  \varphi_{\raisebox{-1pt}{\tiny 3,3  }} 
\vartheta_3(q,z_3) 
 \nn \\
&&  
  - \widetilde{\varphi}_{\raisebox{-1pt}{\tiny 1}}^{\frac{3}{2}}\, \vartheta_3(\mathfrak{q},\mathfrak{z}_{3}) \rgroup \Big ]  
- i \dfrac{ \epsilon}{\widetilde{\Delta}} 
\Big[ x^{\frac{1}{2}}  \mathcal{R}_1   \mathrm{Im}\Big \lgroup  \lo 1+ \frac{x}{2} 
\rc \varphi_{\raisebox{-1pt}{\tiny 1,0 }} \vartheta_4(q,z_1)  -
 \dfrac{f}{2}  \varphi_{\raisebox{-1pt}{\tiny 3,1 }}\vartheta_4(q,z_3) \nn \\
&&     + \frac{x}{2} \widetilde{\varphi}_{\raisebox{-1pt}{\tiny 0}}^{\frac{1}{2}}
\vartheta_4(\mathfrak{q},\mathfrak{z}_{1})  \Big \rgroup + x \mathcal{R}_2 \; \mathrm{Im}\Big \lgroup  \frac{1}{2} \lo  1+ x \rc 
\varphi_{\raisebox{-1pt}{\tiny 2,2 }}\vartheta_4(q,z_2) 
 - \dfrac{ f}{3}  \varphi_{\raisebox{-1pt}{\tiny 4,3 }} \vartheta_4(q,z_4) 
\nn \\
&& + \frac{x}{2}
 \widetilde{\varphi}_{\raisebox{-1pt}{\tiny 1}} \vartheta_4(\mathfrak{q},\mathfrak{z}_{2}) \Big \rgroup  
   + \dfrac{x^{\frac{3}{2}}}{6} \mathcal{R}_3 \; 
 \mathrm{Im}\lgroup \varphi_{\raisebox{-1pt}{\tiny 3,3 }}\vartheta_4(q,z_3) \rgroup + \dfrac{x^2}{24}
\mathcal{R}_4 \mathrm{Im}\lgroup  \varphi_{\raisebox{-1pt}{\tiny 4,6 }} \vartheta_4(q,z_4) 
 \rgroup \Big ] \nn \\
&&   \mp \Big [ x^{\frac{1}{2}} \,  \mathcal{I}_1  \mathrm{Im}\Big \lgroup 
\varphi_{\raisebox{-1pt}{\tiny 1,0 }} \vartheta_3(q,z_1) -
 \dfrac{f}{2}  \varphi_{\raisebox{-1pt}{\tiny 3,1 }} \vartheta_3(q,z_3) \Big \rgroup   -  \frac{x}{2} \, \mathcal{I}_2 \;   \mathrm{Im} \Big \lgroup
   \Big(  \widetilde{\varphi}_{\raisebox{-1pt}{\tiny 1}} - \dfrac{f }{3}  \widetilde{\varphi}_{\raisebox{-1pt}{\tiny 3}}  \Big) 
 \vartheta_3(\mathfrak{q},\mathfrak{z}_{2}) \Big \rgroup \nn \\
&& + \dfrac{x^{\frac{3}{2}}}{6} \,  \mathcal{I}_3 \;  \mathrm{Im}\lgroup  \varphi_{\raisebox{-1pt}{\tiny 3,3 }}  \vartheta_3(q,z_3) \rgroup
-\dfrac{x^2}{24} \, \mathcal{I}_4 \;  \mathrm{Im} \Big \lgroup 
 \widetilde{\varphi}_{\raisebox{-1pt}{\tiny 3}}^2 \, \vartheta_3(\mathfrak{q},\mathfrak{z}_{4})\Big \rgroup \Big ] \nn \\
&& \pm   \dfrac{2 \tilde{\epsilon}^2}{\widetilde{\Delta}^2} 
\Big [    x^{\frac{1}{2}} \, \mathcal{I}_1   \mathrm{Im}\Big \lgroup \lo 1+ x \rc \varphi_{\raisebox{-1pt}{\tiny 1,0}} 
\vartheta_3(q,z_1) +
 \dfrac{3f}{2}   \varphi_{\raisebox{-1pt}{\tiny 3,1 }} \vartheta_3(q,z_3) -
 x \widetilde{\varphi}_{\raisebox{-1pt}{\tiny 0}}^{\frac{1}{2}} \vartheta_3(\mathfrak{q},\mathfrak{z}_{1}) \Big \rgroup  \nn \\
&& - x \, \mathcal{I}_2
 \mathrm{Im}\Big \lgroup x  \varphi_{\raisebox{-1pt}{\tiny 2,2 }}  \vartheta_3(q,z_2) -  
\Big(  \Big ( \frac{1}{2} +x \Big ) \widetilde{\varphi}_{\raisebox{-1pt}{\tiny 1}}  + \dfrac{5 f }{6} \widetilde{\varphi}_{\raisebox{-1pt}{\tiny 3}}  \Big) 
\vartheta_3(\mathfrak{q},\mathfrak{z}_{2}) \Big \rgroup
 \nn \\
&&
    + \dfrac{x^{\frac{3}{2}}}{6}   \, \mathcal{I}_3
\mathrm{Im}\Big \lgroup   \varphi_{\raisebox{-1pt}{\tiny 3,3 }}  \vartheta_3(q,z_3)  \Big  \rgroup 
 -  \dfrac{x^2}{24} \, \mathcal{I}_4 
\mathrm{Im}\Big \lgroup \widetilde{\varphi}_{\raisebox{-1pt}{\tiny 3}}^2 \, 
\vartheta_3(\mathfrak{q},\mathfrak{z}_{4}) \Big \rgroup \Big ]\Big),
\label{xi_theta}
\eea
where the  parameters and phases given above read as follows: 
$f = x |\alpha_{+}|^2$, $\tau = \widetilde{\Delta}\, t$,  $q(\tau) = \exp\left(-\frac{1}{2|\alpha_{+}|^2}-i \frac{x^2 \tau}{4}\right)$,
 $\mathfrak{q} = \exp\left(-\frac{1}{2|\alpha_{+}|^2}\right)$,
 $z_{\jmath} (\tau)=\frac{1}{2} \left(x \left(1+ (1 -  \jmath)\,\frac{x}{4}\right) \tau -i \right)$, 
 $\mathfrak{z}_{\jmath}(\tau) =-\frac{1}{2} \left( \jmath \frac{x^2}{4} \tau  + i\right)$,  
$\varphi_{\jmath, \ell} (\tau)=\exp \left(-i \tau \left (1-\jmath \frac{x}{2} + \ell \frac{x^2}{4} \right)\right)$, 
$\widetilde{\varphi}_{\jmath}(\tau)= \exp \left( i \tau x \left( 1-\jmath \frac{x}{4}\right)\right)$,
$\mathcal{R}_{n}  = \mathrm{Re} \lgroup\widetilde{\alpha}^{(n)}\rgroup$, 
$\mathcal{I}_{n}= \mathrm{Im} \lgroup\widetilde{\alpha}^{(n)}\rgroup$, 
$\widetilde{\alpha}^{(n)} = \alpha_{+}^n e^{-i n \omega t}$. 
The linear combination of theta functions used in (\ref{xi_theta}) is given by 
$\Theta(\tau) = \varphi_{\raisebox{-1pt}{\tiny 0,0}}  \vartheta_4(q,z_{0}) -f(1+x)\varphi_{\raisebox{-1pt}{\tiny 2,0}} \vartheta_4(q,z_{2})  
+  \frac{3}{4} f^2 \varphi_{\raisebox{-1pt}{\tiny 4,2}}  \vartheta_4(q,z_{4}).$
This completes our closed form evaluation of the elements of the reduced density matrix
of the qubit via the Jacobi theta functions. For moderately large coupling $( \lambda/\omega \lesssim 0.20)$ this summation technique agrees well with the series evaluations based on  
(\ref{density_diagonal}, \ref{density_offdiagonal}). Concerning our choice of the bias parameter we note that as we have included terms up to $O(\epsilon^2)$,  
we need to maintain the order of the dimensionless ratio  $\epsilon/{\widetilde{\Delta}} \sim 0.1$.
As the norm $|\alpha_{+}|$  appears in the coefficients of the perturbative terms  as well as in the Gaussian factors, we choose it in the optimum range 
$|\alpha_{+}| \sim 1.5-2$.
These restrictions describe the parametric regime in which our theta functions based analytical evaluations hold. 
The fluctuation 
$\zeta$ of the diagonal qubit matrix element is $O(\epsilon)$, 
and in the low bias regime $\epsilon \ll 
\Delta$ the off-diagonal component overwhelms the fluctuation in the diagonal part: 
$|\zeta| /\left|\xi^{(\pm)}\right| \ll 1$. In Fig. \ref{Off_diagonal_compt} we depict the real and  the imaginary parts of the off-diagonal element of the reduced qubit density matrix. Moreover, as the imaginary part of the off-diagonal matrix element  $\xi^{(\pm)}$ is $O(\epsilon)$, in the $\epsilon \ll \Delta$ limit its real part  gives the major contribution. 

\par

Towards  making a  quantitative comparison between the closed form expression (\ref{xi_theta}) of the density matrix element $\xi^{(+)}$ with the corresponding sum given in (\ref{density_offdiagonal})
 we use full time-evolution data presented in Fig. \ref{Off_diagonal_compt}. For both real and  imaginary parts of $\xi^{(+)}$ we compute  the magnitudes of their  differences between the series evaluation (\ref{density_offdiagonal}) and the corresponding  theta function based result (\ref{xi_theta}). The time-averaged  values of these quantities are given by
$\langle| \Delta\mathrm{Re}(\xi^{(+)})|\rangle_{\mathrm{av}} = 0.01174$ and $\langle| \Delta\mathrm{Im}(\xi^{(+)})|\rangle_{\mathrm{av}} = 0.00376$, respectively. These objects obtained via series evaluation (\ref{density_offdiagonal})  read: 
$\langle| \mathrm{Re}(\xi^{(+)})|\rangle_{\mathrm{av}} = 0.12149$ 
and $\langle|\mathrm{Im}(\xi^{(+)})|\rangle_{\mathrm{av}} = 0.0296$. The order of accuracy for the theta function based evaluations in the parametric regime considered here is, 
therefore, around $90\%$. The  accuracy may be further improved by incorporating higher order terms in the bias parameter $\epsilon$.

\par

The revival and collapse process of $\xi^{(\pm)}$  is evident in Fig. \ref{Off_diagonal_compt}.
To understand the revival time of the off-diagonal element $\xi^{(\pm)}$ we, for simplicity,  study the time dependent terms in (\ref {xi_theta}) for the limiting case $\epsilon = 0$ and $x \ll 1$, where the dominant term  is $O(\sqrt{x})$.
For the range $|\alpha_{+}| \sim 1.5-2$ the structure of the corresponding terms in 
(\ref {xi_theta}) suggests that the oscillation with a \textit{very} short time period 
($T_{\hbox{\tiny{short}}}$) occurs due to the phase  
$\varphi_{1, 0}$, whereas a longer time period ($T_{\hbox{\tiny{revival}}}$) is generated by the leading phase factor in the theta function. Up to the order of approximation described here these periods read
$T_{\hbox{\tiny{short}}} = \pi\big((1 - \frac{x}{2})\, \widetilde{\Delta}\big)^{-1}$ and 
$T_{\hbox{\tiny{revival}}} = \pi \big(x\, {\widetilde{\Delta}})^{-1}$, respectively.
 For $\alpha = 2, \lambda = 0.08 \, \omega$ their numerical values measured in the scaled time read  as $T_{\hbox{\tiny{short}}} = 21.5$ and $T_{\hbox{\tiny{revival}}} = 828.7$, which show good agreement with the corresponding results ($22.9$ and $821.6$) obtained via (\ref{density_offdiagonal}). For a relatively higher coupling strength $\lambda/\omega \lesssim 0.2$ more terms in (\ref {xi_theta}) need to be considered to estimate $T_{\hbox{\tiny{revival}}}$. At a still higher value of $\lambda$ the picture of revival becomes fuzzy as a large number of modes with incommensurate frequencies produce a phase randomization.

\par

\begin{figure}[]
\begin{center}
\includegraphics[width=8cm,height=3.5cm]{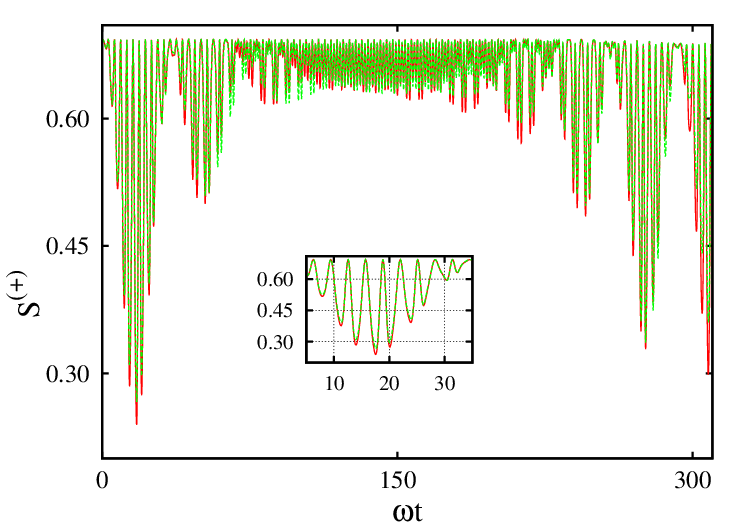}
\end{center}
\caption{Evaluation of the entropy  for   $\lambda=0.15\, \omega , \Delta=0.15\,\omega ,
\epsilon=0.01 \,\omega\;  \mathrm{and} \; \alpha=2 $ 
 using the series (\ref{density_diagonal}, \ref{density_offdiagonal})\, (red solid), and  theta function (\ref{zeta_theta}, \ref{xi_theta})\, (green dashed).   }
\label{entropy_plot}
\end{figure}

The time evolution of the entanglement of the  quasi-Bell initial state (\ref{quasi_Bell})   may be measured by the corresponding von Neumann entropy $S^{(\pm)}$.
Employing (\ref{entropy_defn}) and our evaluations (\ref{zeta_theta}, \ref{xi_theta}) we may find a theta function dependent expression for the time evolution of the  entropy in the parametric domain discussed earlier.  This may be compared with the expression of the entropy based on the sums (\ref{entropy_defn}, \ref{density_diagonal}, \ref{density_offdiagonal}) as a check on the validity of our closed form evaluation scheme. As these expressions are voluminous, we only reproduce  in Fig. \ref{entropy_plot} the corresponding  diagrams for the time evolution of the entropy. Up to the first few revivals   for $\xi^{(+)}$ we observe that the entropy obtained via the series sum is well-approximated by the analytical closed form expression. During the revival of the qubit density matrix elements the entropy of the system  exhibits drastic fall from its near-maximal value 
signaling existence of \textit{almost} disentangled pure states of the \textit{subsystems} at corresponding times during their course of evolution. We will  discuss  in Subsec. \ref{sqeezing_Mandel} an instance of this phenomenon in detail. The extent of the dip in entropy increases with increase of the bias parameter. 

\section{The  $Q$-function of the oscillator density matrix}
\label{Q-function}
\setcounter{equation}{0}
The Husimi $Q$-function [\cite{Schleich2001}] is a quasi-probability distribution  on the complex phase space that provides a description of the time evolution of the
quantum state of the oscillator. It assumes non-negative values and is normalized to unity on the phase space. For our choice of the quasi-Bell initial states 
(\ref{quasi_Bell}) it is defined as the expectation value of  the reduced density matrix of the oscillator (\ref{o_density_matrix}) in an arbitrary coherent state:
\beq
Q^{(\pm)}(\beta)=\dfrac{1}{\pi} \langle \beta | \rho_{\cal O}^{(\pm)}(t)| \beta \rangle.
\label{Q_def}
\eeq
For the state under consideration its explicit positive definite form reads
\bea
Q^{(\pm)}(\beta)=\dfrac{1}{2 \pi} \exp(-|\alpha_{+}|^2) \Big (  \exp(-|\beta_{+}|^2 )| \mathrm{X}^{\pm}|^2 + 
\exp(-|\beta_{-}|^2 )| \mathrm{Y}^{\pm}|^2
 \Big),
\label{Q_factorized}
\eea
where the Fourier sums  are given by
\beq 
\mathrm{X}^{\pm} = \sum_{n=0}^{\infty} \dfrac{(\alpha_{+} \beta_{+}^*)^{n}}{n!}\mathsf{C}_{n}^{\pm}(t)\,\exp (-i n \omega t),\quad
\mathrm{Y}^{\pm} = \sum_{n=0}^{\infty} (-1)^{n} \dfrac{(\alpha_{+} \beta_{-}^*)^{n}}{n!}\mathsf{C}_{n}^{\mp}(t)^{*}\,\exp (-i n \omega t).
\label{XY_def}
\eeq
In  (\ref{Q_factorized}, \ref{XY_def}) we have abbreviated: $\beta_{+}=\beta + \lambda /\omega$ and $\beta_{-}=\beta - \lambda /\omega$. The $Q$-distribution (\ref{Q_factorized}) of the reduced density matrix  (\ref{o_density_matrix}) does not have any zero on the phase space except 
at asymptotically large radial distances. The $Q$-function (\ref{Q_factorized})
 satisfies the normalization condition: $\int Q^{(\pm)}(\beta)\, d^{2}\beta = 1$. For later use we note that, in the $\epsilon=0$ case, the coefficients (\ref{C_biasfree}) enforce the parity constraint:
 \beq
 Q^{(\pm)}(-\beta)|_{\epsilon=0}=Q^{(\pm)}(\beta)|_{\epsilon=0}.
 \label{Q_parity}
 \eeq

\par

\begin{figure}[]
\begin{center}
\subfloat[]{\includegraphics[width=3.3cm,height=3.3cm]{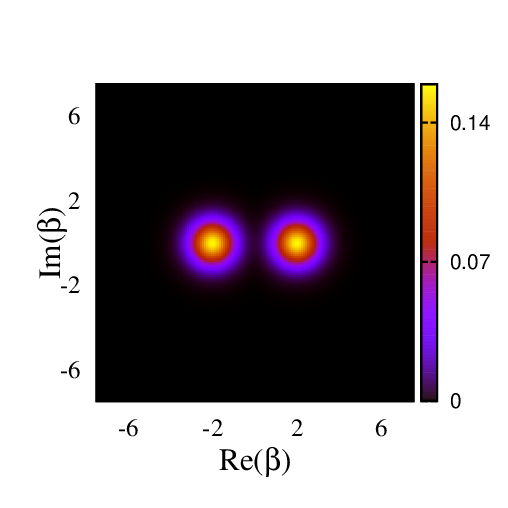}} 
\captionsetup[subfigure]{labelformat=empty}
  \subfloat[(b)]{\includegraphics[width=3.3cm,height=3.3cm]{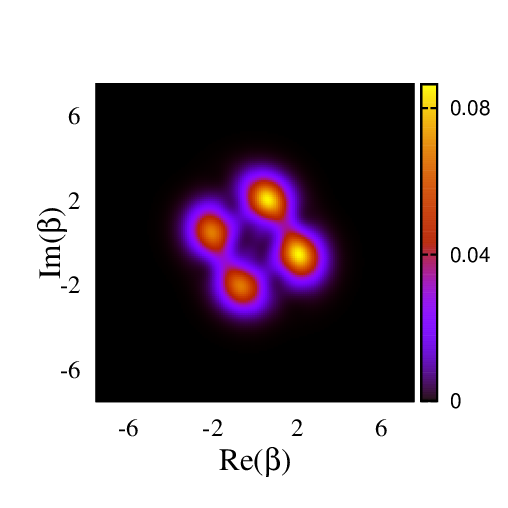}} 
\captionsetup[subfigure]{labelformat=empty}
\subfloat[(c)]{\includegraphics[width=3.3cm,height=3.3cm]{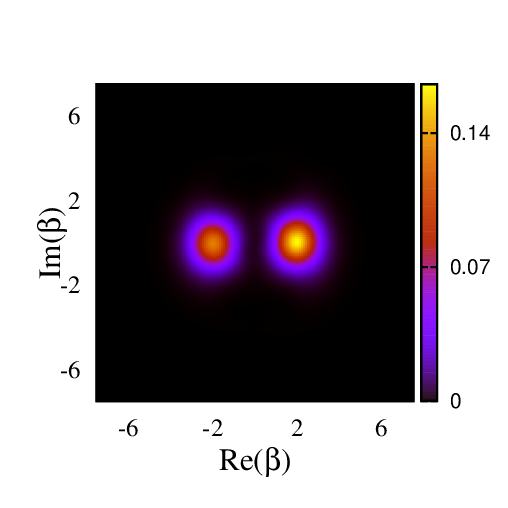}}
\captionsetup[subfigure]{labelformat=empty} 
\end{center}
\caption{The $Q^{(+)}(\beta)$ function using (\ref{XY_def})  for $ \Delta=0.15\, \omega, \epsilon=0.01 \omega$ and $\alpha=2$ at 
various values of scaled time (a) $0$, (b) $155$ and (c) $300$ in 
the strong coupling domain: $\lambda=0.15\, \omega$.} 
\label{qfunc_fig} 
 \end{figure} 
To study the phase space $Q$-distribution (\ref{Q_factorized}) we consider its time-evolution for both the strong (Fig. \ref{qfunc_fig}) and  the ultra-strong coupling regime. 
We distinguish between two cases: ($\mathbf {i}$) For the  strong coupling regime where, say, the parametric values described following (\ref{xi_theta}) hold, the Laguerre polynomials may be
approximated by their quadratic parts.  The off-diagonal element in the Hamiltonian (\ref{H-block}) then contains  effective interaction terms up to $O(x^{2} \widetilde{\Delta})$. In the \textit
{short-term time scale} $t \ll (x^{2} \widetilde{\Delta})^{-1}$  observed in Fig. \ref{qfunc_fig} the corresponding $Q^{(+)}(\beta)$ function exhibits  a quasi-periodic behavior indicating that an interference pattern  develops between the harmonics of the interaction-dependent  mode with frequency $O(x \widetilde{\Delta})$. 
This  produces pronounced standing waves for some time interval  causing bifurcation of each one of the twin quasi-probability peaks, and consequent  spreading of the $Q^{(+)}(\beta)$ distribution in the phase space. Signifying a to and fro transfer of energy between the qubit and the oscillator, the periodic splitting and rejoining of the peaks of the $Q$-distribution correspond, respectively, to the collapse and the revival of the qubit  off-diagonal element $\mathrm{Re}(\xi^{(+)})$ (\ref{density_offdiagonal}). The physical process is reminiscent of an {\sf LC} circuit. The time interval ($T_{\hbox{\tiny{split}}}$) is roughly half of the revival time observed for  
$\mathrm{Re}(\xi^{(+)})$:  $T_{\hbox{\tiny{split}}} \approxeq T_{\hbox{\tiny{revival}}}/2$. This can be readily inferred from Figs. \ref{Off_diagonal_compt} (a) and \ref{qfunc_fig} (b).  Interaction modes of frequency $O(x^{2} \widetilde{\Delta})$ generate quasi-periodic behavior in a \textit{very long time scale}. This will be discussed in detail in Subsec. \ref{Wehrl_entropy} In the presence of a Kerr-type of nonlinear Hamiltonian an  initial coherent state is known to evolve 
[\cite{MTK1990}-\cite{MBWI2001}], at specified times, to discrete superposition of coherent states. We will discuss the similarities and dissimilarities of these results with that of the present example in Subsec. \ref{kitten} 
($\mathbf {ii}$) For the ultra-strong coupling domain $\lambda \sim \,\omega$ 
 interaction modes with frequencies $\{O(x^n \widetilde{\Delta})| n = 0,1, \ldots\}$ and their harmonics rapidly develop and produce a fully randomized interference pattern. The randomness of the phases of the oscillator interaction modes erases any extant phase relation between the ground state and the excited state of the atomic two level system. This causes  disappearance of the collapse and revival properties of the atomic density matrix elements.  The $Q$-distribution  in the phase space now occupies a wider area and exhibits a chaotic evolution pattern without returning to its initial configuration within a finite time. We will discuss these issues in Sec. \ref{delocalization}. 

\par

One of the utilities of the $Q$-distribution is that it provides a convenient evaluation of
the expectation values of the operators expressed in their antinormal ordered form. The
first and second moments of the quadrature variable defined as
\beq
X_{\theta}=\dfrac{1}{\sqrt{2}}(a e^{-i\theta}+a^{\dagger}e^{i\theta})
\label{EB_def}
\eeq
may be expressed [\cite{Schleich2001}] via the following phase space integrals over the complex plane:
\bea
\langle X_{\theta} \rangle^{(\pm)} &\equiv& \hbox {Tr}(X_{\theta}(a,a^{\dagger}) \rho_{\cal O}^{(\pm)}(t)) = \dfrac{1}{\sqrt{2}} \int 
(\beta e^{-i\theta}+ \beta^* e^{i\theta}) Q^{(\pm)}(\beta) d^2 \beta,\nn\\
\langle X_{\theta}^2 \rangle^{(\pm)} &\equiv& \hbox {Tr}(X_{\theta}^2(a,a^{\dagger}) \rho_{\cal O}^{(\pm)}(t)) = \dfrac{1}{2} \int
\Big( (\beta e^{-i\theta}+ \beta^* e^{i\theta})^2-1 \Big ) Q^{(\pm)}(\beta) d^2 \beta.
\label{moment_def}
\eea
Employing the Fourier sum (\ref{Q_factorized}) and using the following notations for the coefficients 
\bea
G_{n;\ell}^{\pm}(\theta;t) &=&\alpha_{+}^{\ell} \; \Big ( \mathsf{C}_{n+\ell}^{\pm}(t){\mathsf{C}_{n}^{\pm}}(t)^* +
{\mathsf{C}_{n+\ell}^{\mp}}(t)^* \mathsf{C}_{n}^{\mp}(t)
 \Big )\; \exp(-i\ell (\omega t + \theta)),\nn\\
 \mathcal{G}_{n;\ell}^{\pm}(\theta;t) &=& \alpha_{+}^{\ell} \; \Big ( \mathsf{C}_{n+\ell}^{\pm}(t){\mathsf{C}_{n}^{\pm}}(t)^* -
{\mathsf{C}_{n+\ell}^{\mp}}(t)^* \mathsf{C}_{n}^{\mp}(t)
 \Big )\; \exp(-i\ell (\omega t+\theta),
 \label{G_Fourier} 
\eea
the first two moments of the quadrature are enlisted as a sum on the  modes:
\bea
\langle X_{\theta} \rangle^{(\pm)} &=& \dfrac{1}{\sqrt{2}} \exp(-|\alpha_{+}|^2)  \sum_{n=0}^{\infty} 
\dfrac{|\alpha_{+}|^{2n}}{n!} 
\mathrm{Re}\Big \lgroup \mathcal{G}_{n;1}^{\pm}(\theta;t)\Big \rgroup  \pm \sqrt{2x}\,\cos \theta \,\zeta,
\label{1st_moment}\\
\langle X_{\theta}^2 \rangle^{(\pm)} &=& \! \! \! |\alpha_{+}|^2 + \dfrac{1}{2} \Big( 1 +x \cos^2 \theta + 
\exp(-|\alpha_{+}|^2)  \sum_{n=0}^{\infty} 
\dfrac{|\alpha_{+}|^{2n}}{n!} \times \nn \\
&&  \times \quad \mathrm{Re}\Big \lgroup G_{n;2}^{\pm}(\theta;t)
-2\sqrt{x} G_{n;1}^{\pm}(\theta;t) \cos \theta \Big \rgroup \! \Big ).
\label{2nd_moment}
\eea
The variance of the quadrature reads:
$V_{\theta}= \langle X_{\theta}^{2} \rangle - \langle X_{\theta} \rangle ^2,$
which we will later employ 
 for studying emergence of the squeezed state during time evolution. 
For later utilization we also quote here the mean photon number $\langle\hat{n}\rangle^{(\pm)} = \hbox{Tr} (a^{\dagger} a\,\rho_{\cal O}^{(\pm)}(t))$ and its variance $\langle \Delta\hat{ n}^{2}\rangle^{(\pm)} = \hbox{Tr} (\hat{n}^{2}\,\rho_{\cal O}^{(\pm)}(t)) - \big(\langle\hat{n}\rangle^{(\pm)}\big)^{2}$: 
\bea
\langle \hat{n}\rangle^{(\pm)}&=& 
|\alpha_{+}|^2+ \frac{\lambda^2}{\omega^2} -\frac{\lambda}{\omega}\exp(-|\alpha_{+}|^2) \sum_{n=0}^{\infty} 
\frac{|\alpha_{+}|^{2n}}{n!} \mathrm{Re}\lgroup G_{n;1}^{\pm}(t)\rgroup,
\label{photon_n1}\\
\langle \Delta\hat{ n}^{2}\rangle^{(\pm)}&=&|\alpha_{+}|^2+ \frac{\lambda^2}{\omega^2}
(1+ 2 |\alpha_{+}|^2)- \frac{\lambda}{\omega}\exp(-|\alpha_{+}|^2) \sum_{n=0}^{\infty} 
\frac{|\alpha_{+}|^{2n}}{n!} \times \nn \\
&& \times (1+2n-2|\alpha_{+}|^{2})  \mathrm{Re}\lgroup G_{n;1}^{\pm}(t)\rgroup -\frac{\lambda^2}{\omega^2} 
 \exp(-2|\alpha_{+}|^2) \times\nn \\
&& \times \Big [\sum_{n=0}^{\infty} \frac{|\alpha_{+}|^{2n}}{n!}
\mathrm{Re}\lgroup G_{n;1}^{\pm}(t)\rgroup \Big]^2  + \frac{\lambda^2}{\omega^2} 
 \exp(-|\alpha_{+}|^2) \sum_{n=0}^{\infty} \frac{|\alpha_{+}|^{2n}}{n!}
\mathrm{Re}\lgroup G_{n;2}^{\pm} (t)\rgroup.
\label{photon_n2}
\eea

\par

\begin{figure}[]
 \begin{center}
  \includegraphics[width=8cm,height=3.5cm]{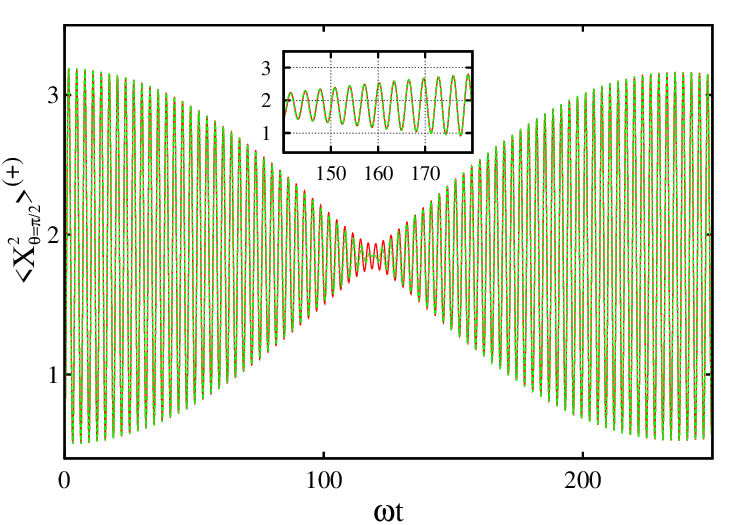}
 \end{center}
\caption{The evaluation of the second moment of the quadrature variable $\langle X^2_{\pi/2} \rangle^{(+)}$ for the values
 $\lambda=0.16\, \omega, \Delta=0.15\, 
\omega ,\epsilon=0.01 \, \omega $ and $ \alpha=1$  using the series (\ref{2nd_moment}) (red solid), and  theta function (\ref{2nd_moment_theta}) (green dashed).  }
\label{Second_moment}
\end{figure}

Using the procedure developed in Sec. \ref{qubit} and the $Q$-function (\ref{Q_factorized}), an arbitrary moment of the oscillator
 dynamical variables may be obtained  as linear combinations of the Jacobi theta functions. We now demonstrate this for 
 the first two moments of the quadrature variable  (\ref{1st_moment}, \ref{2nd_moment}). 
  To this end we retain up to the quadratic terms in the expansion of the Laguerre polynomials and maintain the range of 
 parameters described following 
(\ref{xi_theta}). 
The first moment of the quadrature  is non-vanishing only in the presence of a bias term $(\epsilon \neq 0$) in the qubit Hamiltonian (\ref{Hamiltonian}):
\bea
 \langle X_{\theta} \rangle^{(\pm)} &=&   \dfrac{\epsilon }{ \widetilde{\Delta}}
\dfrac{ \exp(-|\alpha_{+}|^2/2)}{\sqrt{4 \pi |\alpha_{+}|^2}} 
\left(\mathcal{I}_{1}^{\theta} \;
\mathrm{Im} \lgroup \mathsf{T}  \rgroup 
\pm \; \mathcal{R}_{1}^{\theta} \; 
 \mathrm{Re}\lgroup \mathsf{H}\rgroup   \right) \pm \sqrt{2x} \cos \theta \; \zeta,
\label{1st_moment_theta}
\eea
where the linear combinations of the Jacobi theta functions read, respectively, as
\bea
\mathsf{T} &=&  \Big( (2+x+x^2) \sqrt{\widetilde{\varphi}_{\raisebox{-1pt}{\tiny 0}}}+ \Big(2+\dfrac{7x}{2}\Big) f
 \sqrt{\widetilde{\varphi}_{\raisebox{-1pt}{\tiny 2}}}
+\dfrac{3}{2}f^2 \sqrt{\widetilde{\varphi}_{\raisebox{-1pt}{\tiny 4}}} \Big)
  \vartheta_3(\mathfrak{q},\mathfrak{z}_1) \nn \\
&& + x(1+x) \varphi_{\raisebox{-1pt}{\tiny 1,0}}
\vartheta_3(q,z_1) + \dfrac{3}{2} x f \varphi_{\raisebox{-1pt}{\tiny 3,1}} \; \vartheta_3(q,z_3),
\nn\\
\mathsf{H} &=& \Big (x (1+x) \sqrt{\widetilde{\varphi}_{\raisebox{-1pt}{\tiny 0}}}
 -\dfrac{3}{2} x f \sqrt{\widetilde{\varphi}_{\raisebox{-1pt}{\tiny 2}}} \Big)
   \vartheta_4(\mathfrak{q},\mathfrak{z}_1)   \nn \\
&&  -x(1+x) \varphi_{\raisebox{-1pt}{\tiny 1,0}} \vartheta_4(q,z_1) 
+ \dfrac{3}{2} x f \varphi_{\raisebox{-1pt}{\tiny 3,1}}  \vartheta_4(q,z_3),
\label{1st_moment_TH}  
\eea
and  $\mathcal{R}_{n}^{\theta} = 
\mathrm{Re} \lgroup \widetilde{\alpha}^{(n)} \exp(-i n \theta) \rgroup, \;
\mathcal{I}_{n}^{\theta}= \mathrm{Im}  \lgroup \widetilde{\alpha}^{(n)} \exp(-i n \theta) \rgroup$.
Similarly, the second moments of the quadrature   may also be expressed via the Jacobi  theta functions as follows:
\bea
\langle X_{\theta}^2 \rangle^{(\pm)} &=&  \dfrac{1}{2} (1+x \cos^2 \theta)+ |\alpha_{+}|^2 
+ \dfrac{ \exp(-|\alpha_{+}|^2/2)}{\sqrt{2 \pi |\alpha_{+}|^2}}
\times\nn\\ 
& &\times \left(\mathcal{R}_{2}^{\theta} \;
\mathrm{Re}\lgroup \mathcal{T} \rgroup  
\pm \mathcal{I}_{2}^{\theta} \;
\mathrm{Im}\lgroup \mathcal{H} \rgroup\ - 2\sqrt{x} \cos \theta \; \left(\mathcal{R}_{1}^{\theta} \mathrm{Re}\lgroup\mathbf{T}\rgroup 
\mp \mathcal{I}_{1}^{\theta} \mathrm{Im}\lgroup\mathbf{H}\rgroup\right)\right),
\label{2nd_moment_theta}
\eea
where the respective linear sums of $\vartheta_{3}$ and $\vartheta_{4}$ functions read:
\bea
\mathcal{T} &=& \Big( 1-\dfrac{\epsilon^2 x^2}{\widetilde{\Delta}^2}  \Big ) 
\widetilde{\varphi}_{\raisebox{-1pt}{\tiny 1}}  \vartheta_3(\mathfrak{q},\mathfrak{z}_2)
+ \dfrac{ \epsilon^2  x^2 }{\widetilde{\Delta}^2} \varphi_{\raisebox{-1pt}{\tiny 2,1}}  \vartheta_3(q,z_2), \nn\\
\mathcal{H}&=& \Big \lgroup \Big ( 1-\dfrac{\epsilon^2}{2 \widetilde{\Delta}^2} \Big (1+2x+\dfrac{11x^2}{2} \Big)\Big ) 
\widetilde{\varphi}_{\raisebox{-1pt}{\tiny 1}} +\dfrac{ \epsilon ^2}{\widetilde{\Delta}^2} f (1+4x)  
\widetilde{\varphi}_{\raisebox{-1pt}{\tiny 3}} \nn \\ 
&&  - \dfrac{5 \epsilon ^2}{4\widetilde{\Delta}^2} f^2  
\widetilde{\varphi}_{\raisebox{-1pt}{\tiny 5}} \Big \rgroup  \vartheta_4(\mathfrak{q},\mathfrak{z}_2)
-\dfrac{\epsilon^2}{\widetilde{\Delta}^2} x
 \Big (1+\dfrac{11 x}{4}\Big)\varphi_{\raisebox{-1pt}{\tiny 2,1}} \vartheta_4(q,z_2) \nn \\
&& +\dfrac{5 \epsilon^2}{2 \widetilde{\Delta}^2}  x f \varphi_{\raisebox{-1pt}{\tiny 4,3}}
\vartheta_4(q,z_4),\nn\\
\mathbf{T}&=& \Big(1-\frac{\epsilon^2}{\widetilde{\Delta}}(1+x) \Big)\varphi_{\raisebox{-1pt}{\tiny 1,0}}
\vartheta_3(q,z_1) -\frac{2 \epsilon^2}{\widetilde{\Delta}^2}f\varphi_{\raisebox{-1pt}{\tiny 3,1}} 
\vartheta_3(q,z_3) \nn \\
&&
+ \frac{\epsilon^2}{\widetilde{\Delta}^2} 
\Big((1+x) \sqrt{\widetilde{\varphi}_{\raisebox{-1pt}{\tiny 0}}} 
+2 f \sqrt{\widetilde{\varphi}_{\raisebox{-1pt}{\tiny 2}}}\Big ) \vartheta_3(\mathfrak{q},\mathfrak{z_1}) \nn \\
\mathbf{H} &=& \Big(1-\frac{\epsilon^2}{2 \widetilde{\Delta}}(1+x) \Big)\varphi_{\raisebox{-1pt}{\tiny 1,0}}
\vartheta_4(q,z_1)+\frac{ \epsilon^2}{\widetilde{\Delta}^2}f\varphi_{\raisebox{-1pt}{\tiny 3,1}} 
\vartheta_4(q,z_3) \nn \\
&&
-\frac{\epsilon^2}{2 \widetilde{\Delta}} x  \sqrt{\widetilde{\varphi}_{\raisebox{-1pt}{\tiny 0}}}
\vartheta_4(\mathfrak{q},\mathfrak{z_1}).
\label{2nd_moment_theta_comb}
\eea
This completes 
our analytical evaluations of the expectation values of antinormally ordered operators 
in closed forms involving linear combinations of the theta functions. The procedure is valid for the  parametric regime described in Sec. \ref{qubit}. 
In Fig. \ref{Second_moment} we compare the values of the second moment $\langle X^2_{\theta=\frac{\pi}{2}} \rangle^{(+)}$ 
 given by the Fourier sum (\ref{2nd_moment}) and its theta function based analytical evaluation (\ref{2nd_moment_theta}).
 For a quantitative comparison we consider the magnitudes of the  differences between the series evaluation (\ref{2nd_moment})
and the corresponding  theta function based computation (\ref{2nd_moment_theta}).
Its time-averaged  value over the data presented in Fig. \ref{Second_moment} is given by
$\langle|\Delta X_{\pi/2}^2{}^{(+)}|\rangle_{\mathrm{av}} = 0.03886$, whereas the time-averaged series evaluation (\ref{2nd_moment})  reads: $\langle| X_{\pi/2}^2{ }^{(+)}|\rangle_{\mathrm{av}} = 1.84296$. 
 The level of accuracy for the theta function based evaluations in the parametric 
 regime considered here is roughly around $98\%$.

\section{Delocalization at ultra-high coupling strength}
\label{delocalization}
\setcounter{equation}{0}
Our objective in this section is to study the phase space behavior of the system in the ultra-strong
 coupling domain: $\lambda \sim \omega$. Here we \textit{do not} adhere to the  parametric restrictions mentioned following (\ref{xi_theta}) except for the fact that the \textit{adiabatic limit} $\Delta \ll \omega$ [\cite{AN2010}] \textit{still holds}. 
To explore  the evolution (\ref{Q_factorized}) of the phase space density
  $Q^{(\pm)}(\beta)$ we use a number of tools described below. 
  \begin{figure}[H]
\subfloat[]{ \includegraphics[width=5.5cm,height=3cm]{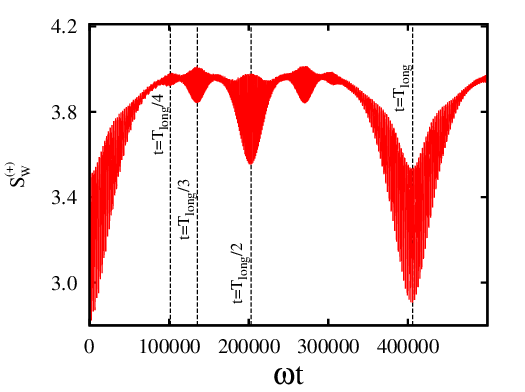}} 
\subfloat[]{\includegraphics[width=5.5cm,height=3cm]{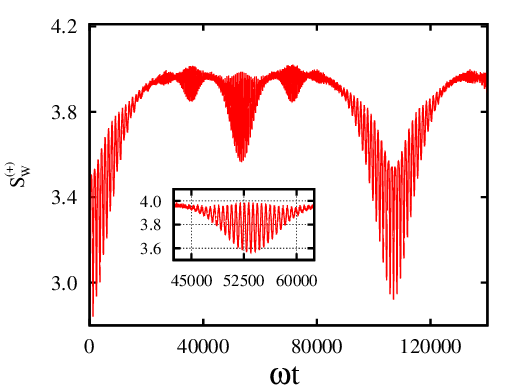}}
 \subfloat[]{\includegraphics[width=5.5cm,height=3cm]{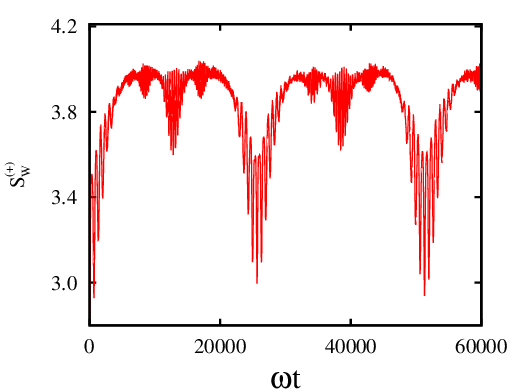}}\\
\subfloat[]{\includegraphics[width=5.5cm,height=3cm]{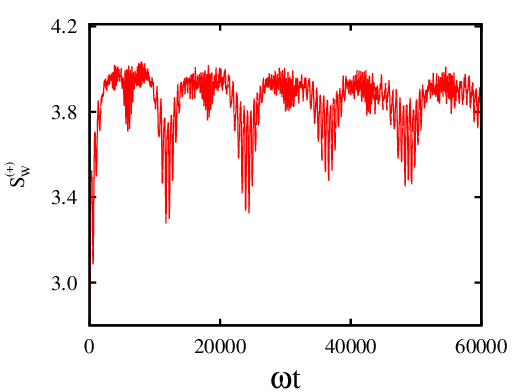}}
\subfloat[]{\includegraphics[width=5.5cm,height=3cm]{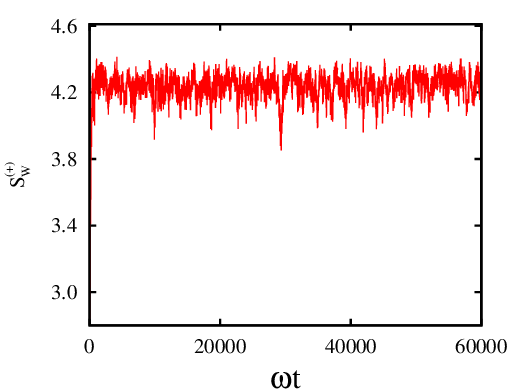}}
\subfloat[]{\includegraphics[width=5.5cm,height=3cm]{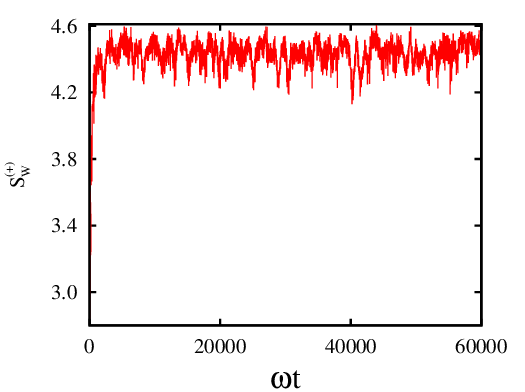}}
\caption{The long time evolution of the Wehrl entropy for  
$ \Delta=0.15 \,\omega, \epsilon=0.03\, \omega$ 
and $\alpha=3$ at various   $\lambda$:
(a)$0.05\,\omega$, (b)$0.07\,\omega$ (c)$0.1\,\omega$, (d)$0.12\,\omega$, (e)$0.9\,\omega$ and 
(f)$1.3\,\omega$.}
\label{Wehrl Entropy}
\end{figure}

  \subsection{Wehrl entropy}  
  \label{Wehrl_entropy}
\begin{figure}
\begin{center}
\includegraphics[width=7.5cm,height=4cm]{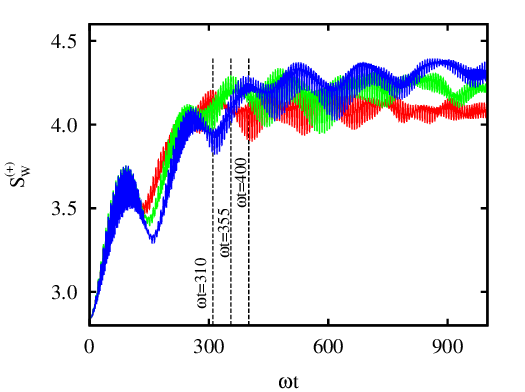}  
\caption{The initial rise of the Wehrl entropy  obtained for 
$ \Delta=0.15 \,\omega, \epsilon=0.03 \,\omega$ 
and $\alpha=3$ at various   $\lambda$=
0.6\,$\omega$ (red),  $\lambda$= 0.8\,$\omega$ (green) and  $\lambda$= 1.0\,$\omega$ (blue).}
\label{initial}
\end{center}
\end{figure}

  The Wehrl entropy [\cite{W1978}] defined as 
\beq
S^{(\pm)}_{W} = - \int  Q^{(\pm)}(\beta)\, \log Q^{(\pm)}(\beta) d^{2}\beta
\label{Wehrl} 
\eeq
 is an information-theoretic measure describing  the delocalization of the oscillator in the phase space. It may be regarded [\cite{BKK1995}] as a count of an equivalent number of widely separated  coherent states necessary for covering the existing phase-space occupation of the coupled oscillator.
It has recently been used for studying quantum phase transitions in several models [\cite{RRC2012}, \cite{CRR2012}]. In the present case we, 
using the time evolution (\ref{Q_factorized}) and the definition (\ref{Wehrl}), 
first numerically study (Fig. \ref{Wehrl Entropy}) the long range time dependence of  Wehrl entropy for various values of coupling strength. We observe the following  properties: ({\bf i}) We first consider the 
 \textit{long range} $t \gtrsim (x^{2} \widetilde{\Delta})^{-1}$ quasi-periodicity of the Wehrl entropy evident in Figs. \ref{Wehrl Entropy} (a, b, c). For the coupling  regime ($\lambda \lessapprox 0.1 \, \omega$) considered therein the Laguerre polynomials in the effective Hamiltonian (\ref{H-block}) are well-approximated by their quadratic components. This generates  interaction-dependent frequency mode $\sim\, x^{2}\, \widetilde{\Delta}$ and its harmonics. Their interference gives rise to quasi-periodic long time  behavior of the coupled oscillator, where the time period satisfies the property $T_{\tiny\hbox {long}} \propto \lambda^{- 4}\,\exp(x/2)$. The time period observed in  
 Figs. \ref{Wehrl Entropy} (a, b, c) are given in our scale as 
$T_{\hbox{\tiny{long}}} = 406081.5\; (\hbox{for}\,\lambda = 0.05\, \omega), T_{\hbox{\tiny{long}}} = 106976.7\; (\hbox{for}\, \lambda = 0.07\, \omega), T_{\hbox{\tiny{long}}} = 25656.6\; (\hbox{for}\, \lambda = 0.10\, \omega)$, respectively. The near equality of the quantity
$\lambda^{4}\, T_{\hbox{\tiny{long}}} \exp(-x/2)$ in the respective cases: $2.52535, 2.54346, 2.51486$ is a check on the validity of the said truncation of the Laguerre polynomials. This explains the long timespan quasi-periodicity of the system in the coupling regime $\lambda \lessapprox 0.1 \, \omega$. We note that a similar behavior in the time-evolution of Wehrl entropy was observed [\cite{JO1994}] in a Kerr-like medium. The local minimum in the said evolution diagram 
[\cite{JO1994}] corresponded  with formation of finite superposition of coherent states. The \textit{long range behavior}  (Figs. \ref{Wehrl Entropy} (a, b, c)) of the Wehrl entropy 
in the present
model qualitatively agrees with the properties obtained in  [\cite{JO1994}]. However, the 
qubit-oscillator interactions in our case induces undulations of frequency $O(x \widetilde{\Delta})$ superposed on the long range oscillations, which now forms an envelop 
(inset of  Fig. \ref{Wehrl Entropy} (b)) of the total time-evolution of the Wehrl entropy. The physical effects of multiple time scales present in our case will be discussed in Subsec. \ref{kitten} ({\bf ii}) We  now comment on the initial rate of production of Wehrl entropy in the ultra-strong coupling domain 
$x \sim 1 \Rightarrow \lambda \gtrsim \omega/2$.  With increased qubit-oscillator coupling  all high-frequency quantum fluctuation modes are now rapidly generated. Interference among these incommensurate modes cause, in general, quick spread the $Q$-distribution in the  phase space resulting in fast initial production of Wehrl entropy $S_{W}^{(\pm)}$. The typical time scale for establishing  the interference pattern  is given, via  (\ref{H-block}), by 
$T_{\hbox{\tiny{ent. prod.}}}\, \widetilde{\Delta}\,L_n(x) \sim 1$. To study the coupling-dependence of initial entropy production time scale $T_{\hbox{\tiny{ent. prod.}}}$ we note the behavior  [\cite{BB2008}] of the Laguerre function $L_n(x>0)$ for an asymptotically large value of the index $n$:
\beq
L_n(x) = \frac{n^{-\frac{1}{4}}}{\sqrt{\pi}} \frac{\exp(x/2)}{x^{\frac{1}{4}}} \cos\left(2 \sqrt{nx}- \pi/4 \right) + O\left(n^{-\frac{3}{4}}\right),
\label{Lagurre_asymptotic}
\eeq
which immediately produces the estimate 
 $T_{\hbox{\tiny{ent. prod.}}} \propto \sqrt{\lambda}$. 
To test the above estimate, we  study (Fig. \ref{initial}) the transient initial production time for 
$S_{W}^{(+)}$. Assuming  $T_{\hbox{\tiny{ent. prod.}}} \approx 310$ units for 
the coupling $\lambda = 0.6 \, \omega$, our asymptotic estimate immediately 
allows us to predict the said 
entropy production time for the cases  $\lambda = 0.8 \, \omega$ and $\lambda = 1.0 \, \omega$ as $357.96$ and $400.21$ units, respectively. These predictions are very close to the corresponding  numerical values marked in Fig. \ref{initial}. ({\bf iii})  In the ultra-strong coupling domain $\lambda \sim  \,\omega$   \textit{all} frequency modes $\{O(x^n \widetilde{\Delta})| n = 0,1, \ldots\} $ and their harmonics quickly arise, and due to randomness of  their incommensurate  phases the resultant interference contributions have a tendency to average out. 
\textit{After} the period of rapid initial increment $(t \gg T_{\hbox{\tiny{ent. prod.}}})$, the long time description  of the Wehrl entropy  given in Figs. \ref{Wehrl Entropy} (e, f) suggests that high-frequency ($O(\omega)$) quantum fluctuations are superimposed on an almost stationary value.  The relative magnitude  of the  quantum oscillations in the Wehrl entropy remains small: $\frac{|\Delta S_{W}|}{S_{W}}\Big|_{t \gg T_{\hbox{\tiny{ent. prod.}}}} \ll 1$.  Thus the randomization of the phases of a large number of  modes induces an \textit{effective stabilization} of the occupation in the phase space after an initial build-up. The hierarchy of the frequency scales  present in our model: $O(\widetilde{\Delta}) \ll \omega$ suggests a natural approach towards making the said quasi-stationary property more evident. We qualitatively describe it below. Following [\cite{JJ2007}] we may obtain physically observed quantities after a suitable coarse graining of the time scale that acts as a low-pass filter effectively eliminating the high frequency components in quantum oscillations.  The long time behavior of physical quantities, say $S_{W}^{(+)}$, is set by the interaction modes typically having frequencies $O(\widetilde{\Delta})$, while  
 the high oscillator frequency $\omega$ gives rise to the rapid fluctuations. A coarse graining \textit {\`{a} la} [\cite{JJ2007}] has a smearing effect on the high frequency fluctuations as the implicit time-averaging process described therein has a suitable  resolution time  
$(\widetilde{\Delta})^{-1} \gg T_{\substack{\hbox{\tiny{coarse}}\\ 
 \hbox{\tiny{graining}}}} \gg \omega^{-1}$. 
The contrast between the long time evolution in the strong coupling (Fig. \ref{Wehrl Entropy} (a, b, c)) and the ultra-strong coupling
 (Figs. \ref{Wehrl Entropy} (e, f))
regimes may be stated as follows. In the former case the system undergoes an almost periodic dynamical evolution, while in the latter case  the initial state does not recur  within a long span of time wherein a quasi-stationary state prevails. This points towards the emergence of a statistical equilibrium. The changeover between the two regimes is evident in Fig. \ref{Wehrl Entropy} (d).  Another significant fact in Fig. \ref{Wehrl Entropy} is that for a fixed value of the phase-space separation $\alpha$ and  in the
 $\lambda\sim \omega$ regime, the time-averaged value of the Wehrl entropy  gradually \textit{increases} with increasing coupling strength. 
The averaging is considered \textit{after} the transient period of entropy production is over, 
and  the time scale of averaging is taken to be much \textit{longer} than $\omega^{-1}$.
 An increasing expectation value of the number operator $\langle\hat{n}\rangle$ due to a stronger 
qubit-oscillator interaction amounts to a wider spread of the $Q$-function in the phase space causing  an increment in the  Wehrl entropy.
 
 \subsection{Kitten states  via long and short time oscillations}
 \label{kitten}
 It is known that in the context of the Kerr-type of nonlinear self-interacting photonic models the local minima in the time evolution of the Wehrl entropy of an initial coherent state are associated  [\cite{JO1994}] with formations of 
 uniformly separated macroscopic superposition of finite number of coherent states 
 [\cite{MTK1990}-\cite{MBWI2001}]. These authors noted that  these transient  superposition of the macroscopic coherent states are realized at rational multiples of the time period of the $Q$-distribution. The number of distinguishable states participating in the superposition depends linearly on the amplitude of the initial coherent state. Experimental realization of a nonclassical superposition of multiple coherent states in a Kerr medium has  been achieved [\cite{Kirchmair2013}] recently. Generation of  the cat-like states in an arbitrary 
finite dimensional bosonic system that admits applying the displacement operation 
on its ground state  has also been studied [\cite{MPPN2014}] in a Kerr-type medium. 
\begin{figure}[H]
\begin{center}
\captionsetup[subfigure]{labelformat=empty}
\subfloat[(a)]{\includegraphics[width=3.2cm,height=3.2cm]{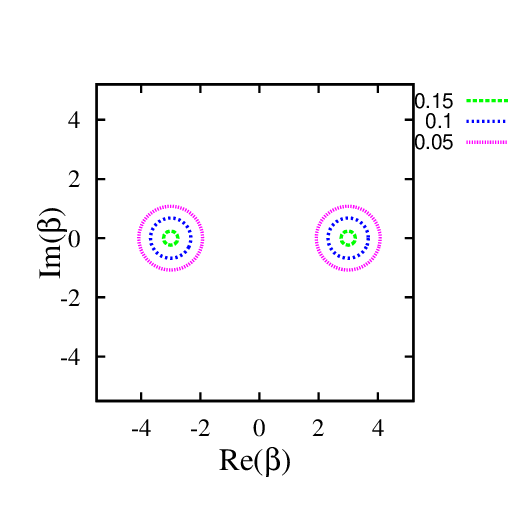}} \hspace{-0.3cm}
\captionsetup[subfigure]{labelformat=empty}
\subfloat[(b)]{\includegraphics[width=3.2cm,height=3.2cm]{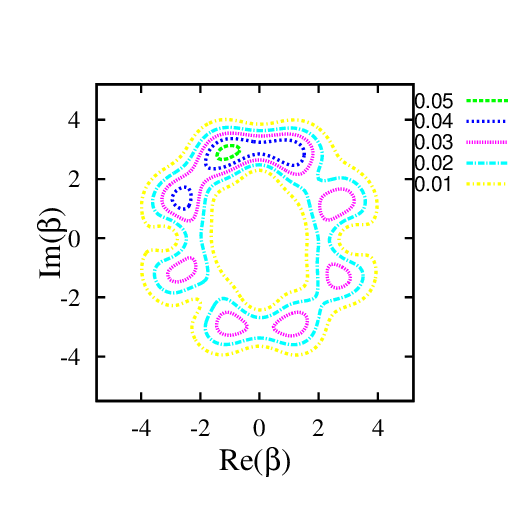}} \hspace{-0.3cm}
\captionsetup[subfigure]{labelformat=empty}
  \subfloat[(c)]{\includegraphics[width=3.2cm,height=3.2cm]{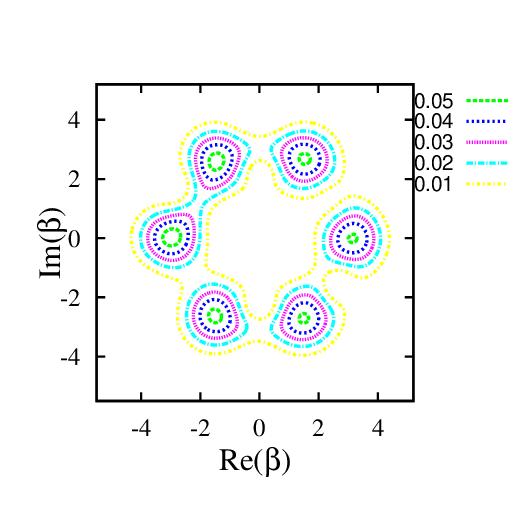}} \hspace{-0.3cm}
\captionsetup[subfigure]{labelformat=empty}
\subfloat[(d)]{\includegraphics[width=3.2cm,height=3.2cm]{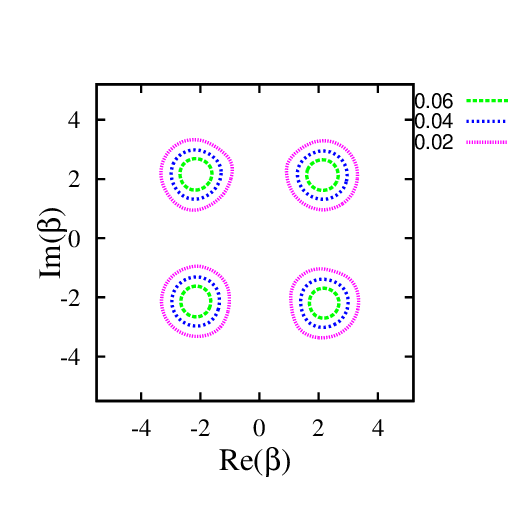}}\hspace{-0.3cm}
\captionsetup[subfigure]{labelformat=empty} 
\subfloat[(e)]{\includegraphics[width=3.2cm,height=3.2cm]{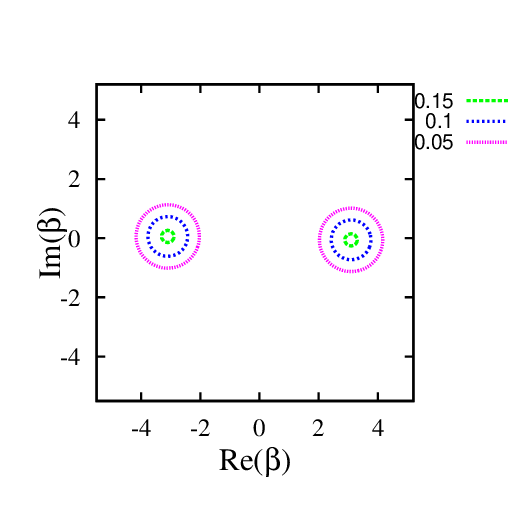}}
\captionsetup[subfigure]{labelformat=empty}
\end{center}
\vspace{-0.5cm}
\caption{The $Q^{(+)}(\beta)$ function using (\ref{XY_def})  for $ \Delta=0.15\, \omega, \epsilon=0.03\, \omega$ and $\alpha=3$ at 
various values of scaled time: (a) $0$, (b)$T_{\mathrm{long}}/4$, (c) $T_{\mathrm{long}}/3$, (d) $T_{\mathrm{long}}/2$, (e) $T_{\mathrm{long}}\,(= 406081.5)$ 
for the chosen coupling strength  $\lambda=0.05\, \omega$.} 
\label{qfunc_fig_quasi} 
\vspace{-0.5cm}
 \end{figure} 

\par

\begin{figure}
\begin{center}
 \subfloat[]{\includegraphics[width=5cm,height=3.1cm]{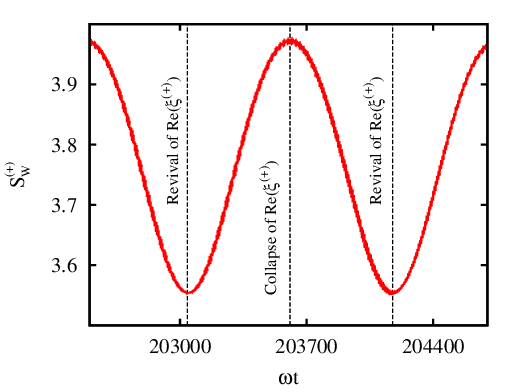}}
\subfloat[]{ \includegraphics[scale=0.4]{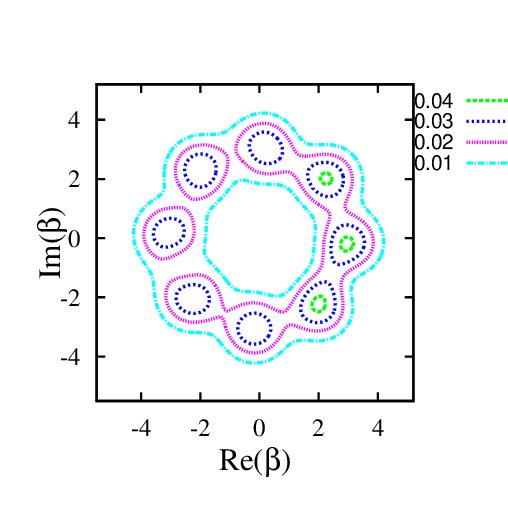} }
\subfloat[]{ \includegraphics[scale=0.4]{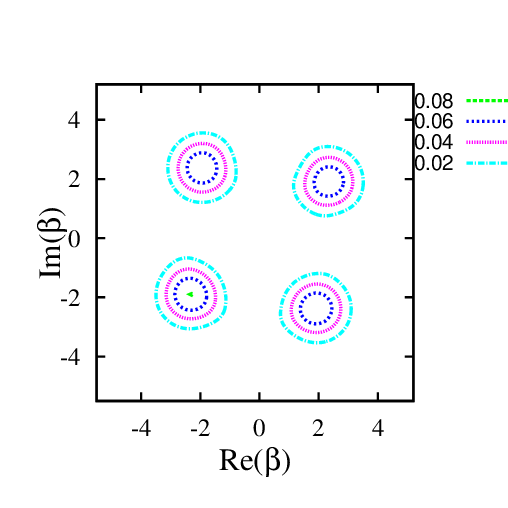} } \\
  \subfloat[]{\includegraphics[width=5cm,height=3.1cm]{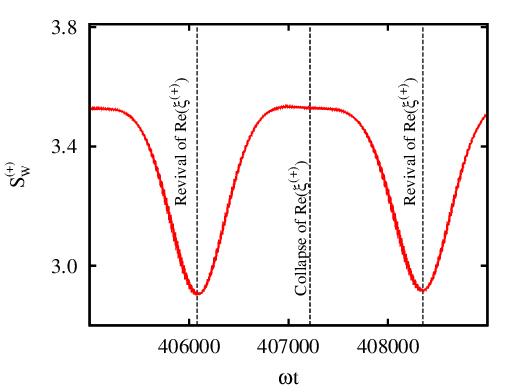}} 
\subfloat[]{\includegraphics[scale=0.4]{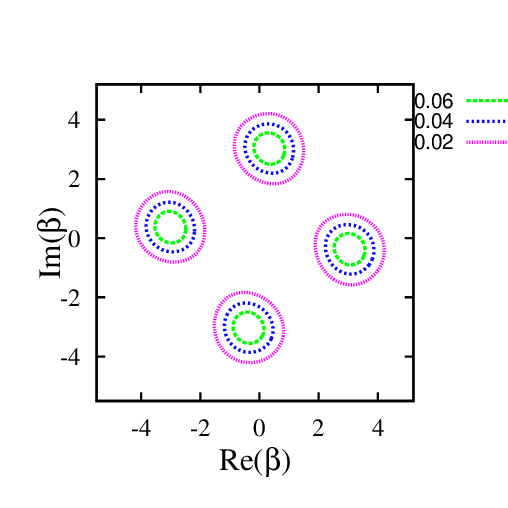}}
\subfloat[]{ \includegraphics[scale=0.4]{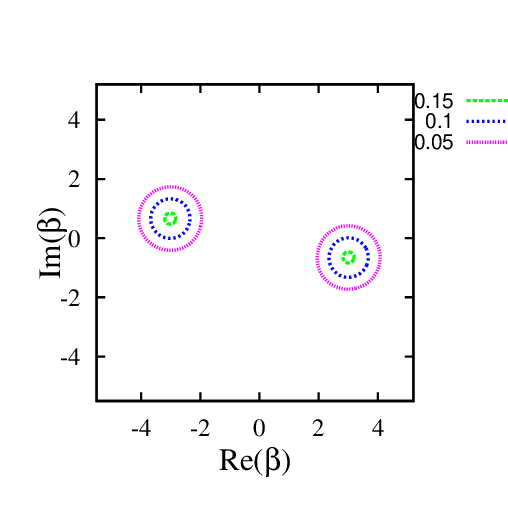}}
 \end{center}
\caption{The evolution of Wehrl entropy is given around (a) $T_{\mathrm{long}}/2$ and (d) $T_{\mathrm{long}}$ for the 
coupling strength $\lambda=0.05\, \omega$, and  $ \Delta=0.15\, \omega, \epsilon=0.03 \omega$ and $\alpha=3$. 
Corresponding to the  local maxima and minima of the Wehrl entropy the $Q$-function is given   at following scaled times: (b) $T_{\mathrm{long}}/2+T_{\mathrm{split}}(T_{\mathrm{long}}/2)$, 
(c) $T_{\mathrm{long}}/2+2\,T_{\mathrm{split}}(T_{\mathrm{long}}/2)$, 
(e) $T_{\mathrm{long}}+T_{\mathrm{split}}(T_{\mathrm{long}})$,
 (f) $T_{\mathrm{long}}+2\,T_{\mathrm{split}}(T_{\mathrm{long}})$, where  
 $T_{\mathrm{long}} = 406081.5, T_{\mathrm{split}}(T_{\mathrm{long}}/2) =567.5 ,T_{\mathrm{split}}(T_{\mathrm{long}}) =1135 $.} 
\label{bifurcation}
 \end{figure}

Much parallel to the scenario [\cite{MTK1990}-\cite{MBWI2001}] in the Kerr medium, our mixed state $Q$-distribution (\ref{Q_factorized})  in the strong coupling limit $\lambda \sim 0.1\, \omega$ represents momentary realizations of macroscopic coherent states  uniformly separated on the phase space  at specific times $T_{p, q} = (p/q)\,T_{\hbox{\tiny{long}}}$, where $(p, q)$ are coprime integers $(p \le q)$. In the small-bias case the parity symmetry (\ref{Q_parity}) of the $Q$-function (\ref{Q_factorized}) requires appearance of an \textit{even}  number ($2 q$) of macroscopically distinguishable Gaussian peaks. This is evident in Fig. \ref {qfunc_fig_quasi}, where the chosen qubit-oscillator coupling $\lambda=0.05\, \omega$ allows the Laguerre polynomials to be approximated by their quadratic components, which produce  modes of frequency $\sim\, x^{2}\, \widetilde{\Delta}$ and its harmonics.
 Starting with the initial quasi-Bell state 
 (\ref{quasi_Bell}) of the composite system, the evolution (Figs. \ref{Wehrl Entropy}
 (a) and \ref{qfunc_fig_quasi}) of the $Q$-function indicates the transient appearances of $8, 6, 4$ uniformly separated Gaussian peaks 
representing macroscopic coherent states, at rational fractions of the long time quasi-period of the Wehrl entropy: $T_{\mathrm{long}}/4, T_{\mathrm{long}}/3, T_{\mathrm{long}}/2$, respectively. However, the presence of \textit{multiple} time scales due to the qubit-oscillator interaction in the present model introduces another novel interference related feature. As explained in the context of 
Fig. \ref{qfunc_fig} (b), it follows that in the vicinity of time $\{T_{p, q} = (p/q)\,T_{\hbox{\tiny{long}}}| (p, q)=1\}$ 
 interference pattern  develops between the harmonics of the interaction-dependent \textit{linear} mode with frequency $O(x \widetilde{\Delta})$. 
This causes an energy transfer in a \textit{short} time scale between the qubit and the oscillator degree of freedom causing a bifurcation (\textit{\`{a} la} Fig. \ref{qfunc_fig} (b)) of the $2 q$ quasi-probability peaks 
to $4 q$ equally separated Gaussian peaks representing coherent state density matrices. 
We illustrate this in Fig. \ref{bifurcation}. The \textit{local} 
variations of Wehrl entropy in the neighborhood of $(p/q)\,T_{\mathrm{long}}$  are given in Figs. \ref{bifurcation} (a) and (d), where we fix $p =1; q=2, 1$, respectively. The Wehrl entropy reaches its local minima
at the stipulated times $T_{\mathrm{long}}/2, T_{\mathrm{long}}$. The corresponding `kitten' states  
with $4$ and $2$ Gaussian peaks are observed in Figs. \ref{qfunc_fig_quasi} (d) and (e), respectively. These regions correspond to the revival of the qubit off-diagonal matrix element
$\mathrm{Re}(\xi^{(+)})$ given in (\ref{density_offdiagonal}). Doubling  of the number of Gaussian peaks to  $4 q$, as it is manifest in Figs. \ref{bifurcation} (b) and (e), occurs at $(1/q)\,T_{\mathrm{long}} + T_{\mathrm{split}}(T_{\mathrm{long}}/q)$ ($q=2,1,$ respectively) coinciding with the \textit{collapse} of the element $\mathrm{Re}(\xi^{(+)})$ that leads to  local \textit{maximization} of the Wehrl entropy. At subsequent revival of the element 
 $\mathrm{Re}(\xi^{(+)})$ at$(1/q)\,T_{\mathrm{long}} +2\, T_{\mathrm{split}}(T_{\mathrm{long}}/q)$ the number of macroscopic Gaussian peaks again reduces to $2 q$. Moreover, the short range splitting time at a particular rational fraction $(p/q)\,T_{\mathrm{long}}$ scales inversely with $q$. For instance, from Figs. \ref{bifurcation} (a) and (d) we observe that 
 $T_{\mathrm{split}} (T_{\mathrm{long}}/2) \approx 
 (1/2) \,T_{\mathrm{split}} (T_{\mathrm{long}})$. In other words, due to the complex nature of the qubit-oscillator interaction in the strong coupling regime $\lambda \sim 0.1\, \omega$ the high frequency quantum oscillations 
 $O(x \widetilde{\Delta})$ are \textit{frequency modulated} by the low frequency component 
 $O(x^{2} \widetilde{\Delta})$. Lastly, a finite bias $\epsilon$ breaks the parity symmetry (\ref{Q_parity}) of the $Q$-function (\ref{Q_factorized}), which may now evolve into an odd number of Gaussian peaks at suitable times.  We will discuss this issue elsewhere.
 
\subsection{Complexity}
\label{complexity_comparison}
Towards understanding the approach to ergodicity in the phase space the authors of Ref. [\cite{SA2002}] introduced another measure of the delocalization that qualitatively reciprocates the Wehrl entropy, which
 is found to be difficult for analytical computation  due to the presence of the logarithmic function in it. They proposed [\cite{SA2002}] the inverse of the second moment of the $Q$-distribution as a measure of complexity of the corresponding quantum states:
\beq
{\cal W}_{2}(Q) = \left({\cal M}_{2}(Q)\right)^{-1},\;  {\cal M}_{2}\big(Q^{(\pm)}\big) = \int \big(Q^{(\pm)}(\beta)\big)^{2}
\, d^{2}\beta. 
\label{complexity_defn}
\eeq 
\begin{figure}[H]
\subfloat[]{ \includegraphics[width=5.5cm,height=3cm]{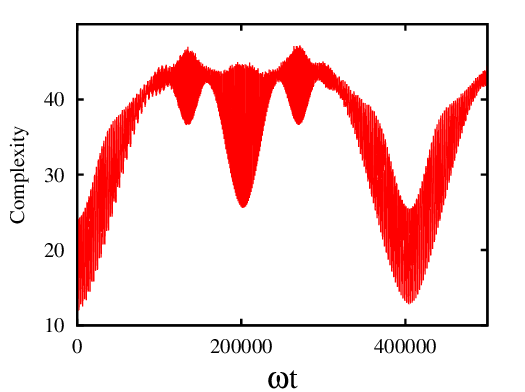}}
\subfloat[]{ \includegraphics[width=5.5cm,height=3cm]{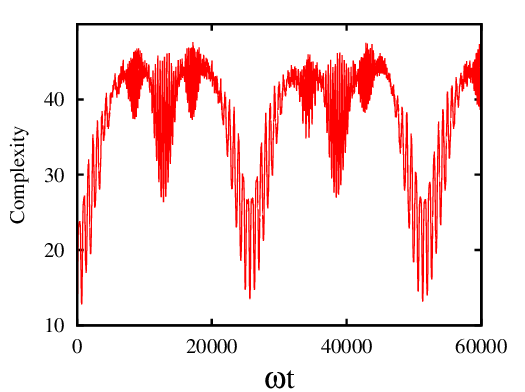}} 
\subfloat[]{\includegraphics[width=5.5cm,height=3cm]{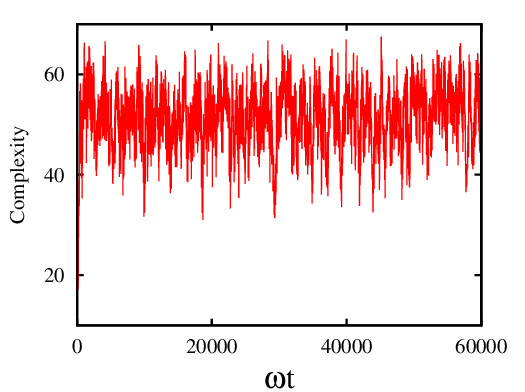}} 
\caption{The long time evolution of the complexity ${\cal W}_{2}(Q)$  obtained  for $ \Delta=0.15\, \omega, \epsilon=0.03\, \omega$ 
and $\alpha=3$ at the following values of   $\lambda$:
(a) 0.05\,$\omega$, (b) 0.1\,$\omega$ and (c) 0.9\,$\omega$.}
\label{Complexity_longtime}
\end{figure}
The complexity of the quantum state ${\cal W}_{2}(Q)$ represents the effective phase space occupied by the $Q$-distribution of the oscillator density matrix. The expansion (\ref{Q_factorized}) allows
us to evaluate the second moment of the $Q$-distribution as follows:
\bea
{\cal M}_{2}\big(Q^{(\pm)}\big) &=& \dfrac{1}{8 \pi} \exp(- 2 |\alpha_{+}|^2) \left\lgroup
\sum_{N =0}^{\infty} 
  \dfrac{|\alpha_{+}|^{2 N}}{2^{N}\,N!} \,  \mathcal {D}_{N}(t) 
  + 2 \exp\left(-\frac{x}{2}\right) \sum_{N,M=0}^{\infty} \alpha_{+}^{N} \big(\alpha_{+}^{*}\big)^{M}\;\;\times\right.\nn\\
& &\left.\times \exp\big(-i (N-M) \,\omega t\big)
\sum_{\mu = 0}^{\min(N, M)}\dfrac{\left(\frac{\lambda}{\omega}\right)^{N + M - 2 \mu}}
{2^{\mu}\,\mu!\,(N-\mu)!\,(M-\mu)!}\qquad\qquad\times\right.\nn\\
 & &\left. \times \mathcal{F}^{\pm}_{N, \mu}(t) \mathcal{F}^{\pm}_{M, \mu}(t)^{*}\right\rgroup, \label{M2}
\eea
where the mode sums on the oscillator states are given by
\bea
\mathcal {D}_{N}(t) &=& \left|\sum_{n = 0}^{N} \binom {N}{n} \mathsf{C}_n^{+}(t)\,
\mathsf{C}_{N-n}^{+}(t)\right|^2 +
\left|\sum_{n = 0}^{N} \binom {N}{n} \mathsf{C}_n^{-}(t)\,
\mathsf{C}_{N-n}^{-}(t)\right|^2,\nn\\
\mathcal{F}^{\pm}_{N, \mu}(t) &=& \sum_{n=0}^{N} \Lambda_{N, n, \mu}\,
\mathsf{C}_n^{\pm}(t)\,\mathsf{C}_{N-n}^{\mp}(t)^{*},\; 
\Lambda_{N, n, \mu} = \sum_{k} (-1)^{k} \binom{\mu}{k}\,\binom{N - \mu}{N - n - k}.
\label{M2_par}
\eea
\label{complexity}
The above expression constitute our analytical evaluation of the complexity 
measure ${\cal W}_{2}(Q)$. 
Our plot for the long time behavior of the complexity ${\cal W}_{2}(Q)$ is given in Fig. \ref{Complexity_longtime}. Agreeing with the assertion in [\cite{SA2002}], its properties faithfully 
resemble those of the Wehrl entropy. In Figs. \ref{Complexity_longtime} (a, b), where the coupling strength remains $\lambda \lesssim 0.1\, \omega$, we find that ${\cal W}_{2}(Q)$ exhibits, exactly parallel to the pattern observed for the Wehrl entropy, a quasi-periodic behavior with the time period inversely proportional to 
$\lambda^{4}\, \exp(-x/2)$. Chaotic character of ${\cal W}_{2}(Q)$ sets in 
(Fig. \ref{Complexity_longtime} (c)) for the ultra-strong coupling regime $\lambda \gtrsim 1.0 \,\omega$.
The initial time  period of a rapid increase in ${\cal W}_{2}(Q)$  in the vicinity of $\lambda \sim 1.0\, \omega$ varies as $\sqrt{\lambda}$. 
We observe in Fig. \ref{Complexity_longtime} (c) that, in the  ultra-high coupling regime $\lambda \gtrsim \omega$ the complexity ${\cal W}_{2}(Q)$ assumes, mimicking the Wehrl entropy, a quasi-stationary state, except for the high frequency random oscillations that may be eliminated by the coarse graining process.
  
\subsection{Quadrature squeezing and Mandel parameter}
\label{sqeezing_Mandel}
\begin{figure}[H]
 \begin{center}
  \subfloat[]{\includegraphics[width=7cm,height=4cm]{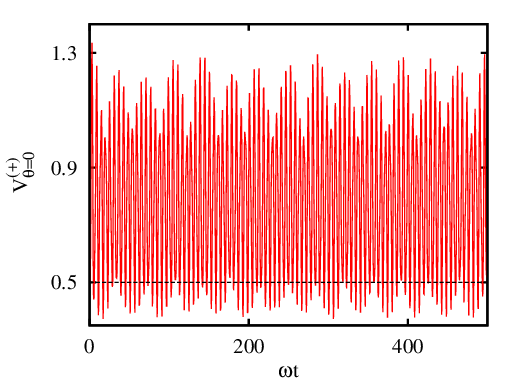}} \quad
\subfloat[]{\includegraphics[scale=0.45]{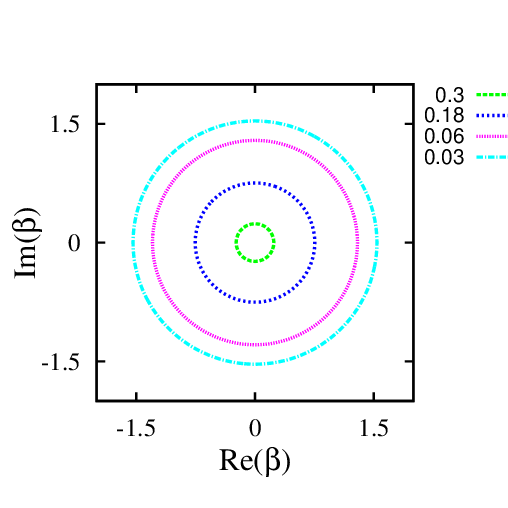}} \quad
\subfloat[]{\includegraphics[scale=0.45]{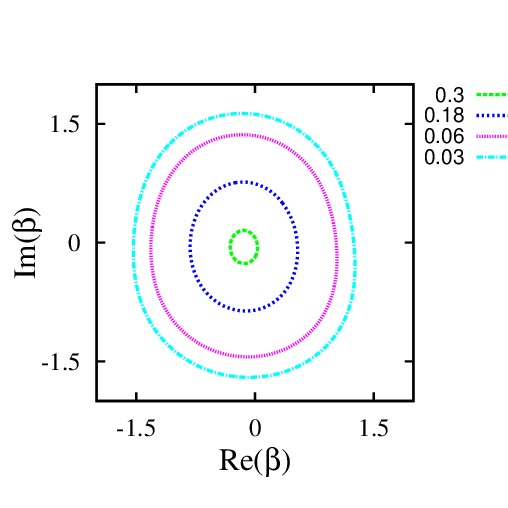}}
\caption{(a) The  time evolution of the quadrature variance $V_{\theta = 0}^{(+)}$, and the $Q$ function at two different scaled times: 
(b) $0$, and (c) $408.5$
for  values $\alpha=0.05$, $\lambda=0.3\, \omega, \Delta= 0.15 \,\omega$ and $\epsilon=0.03 \,\omega$. 
The squeezing at scaled time $\omega t=408.5$  is evident from the corresponding variance $V_{\theta = 0}^{(+)} = 0.3777$.
The contour plot (c) also turns elliptic at the said time.}
\label{squeezed_state}
 \end{center}
\end{figure}

\begin{figure}[H]
 \begin{center}
  \includegraphics[scale=0.5]{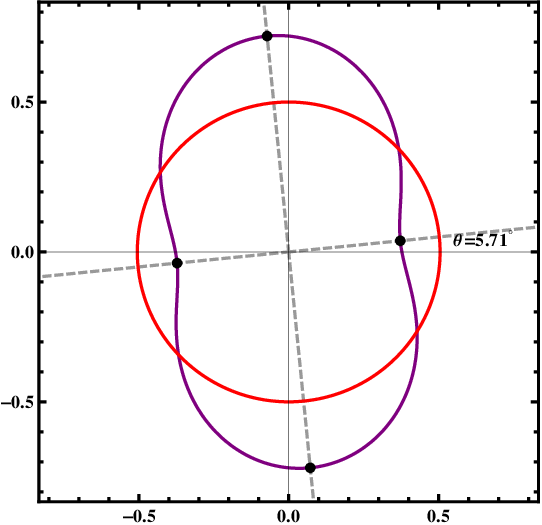}
\caption{The polar plot (w.r.t $\theta$) of the quadrature variance $V_{\theta}^{(+)}$    for
  values $\alpha=0.05$, $\lambda=0.3\, \omega, \Delta= 0.15 \,\omega$ and $\epsilon=0.03 \,\omega$ at time $\omega t=0 $ (red),
and $\omega t =408.5$ (purple) when the least value of  variance $0.3741$ occurs at an angle $\theta=5.71^{\circ}$.}
\label{polar-plot}
 \end{center}
\vspace{-0.4cm}
\end{figure}

For small values of the phase space separation $\alpha$, quantum interferences manifest in the time evolution of the $Q$-distribution (\ref{Q_factorized}) give rise to 
various features of the nonclassicality 
such as  appearance of  squeezed  states and negative values of the  Mandel parameter. The signature of the squeezing is observed in Fig. \ref{squeezed_state} at certain times when the variance $V_{\theta}^{(+)}$ of the quadrature variable,
 say  at $\theta=0$, is rendered less than its classical value $\frac{1}{2}$ that holds for the  normalization (\ref{EB_def}). 
For a small separation $\alpha$ in phase space, the states 
$|\pm \alpha\rangle$ interfere and a contour plot of the $Q$-distribution 
(Fig. \ref{squeezed_state} (c)) makes the quadrature squeezing  evident.  Beginning with a circular shape  (Fig. \ref{squeezed_state} (b)) the $Q$-function turns into an ellipse 
 (Fig. \ref{squeezed_state} (c)) and squeezing appears in an appropriate direction. It follows from Fig. \ref{polar-plot} that at the scaled time $\omega t=408.5$
 the quadrature variance reaches a minimum value $0.3741$ at an angle $\theta=5.71^{\circ}$. The  polar angle  at which the minimum quadrature variance is realized varies with time and depends on the dynamical state of interference of the quantum modes.  An increase in the coupling $\lambda$ first enhances the squeezing as the quadratic term appearing in the expansion of the Laguerre polynomial in the interaction Hamiltonian (\ref{H-block}) plays a key role in producing the squeezing, much parallel to that of a Kerr-like interaction [\cite{TMK1991}].  The squeezing is, however, present during  part of the oscillations. The coarse-graining process [\cite{JJ2007}] discussed before removes the squeezing property as the measured quantity is averaged over time $T_{\substack{\hbox{\tiny{coarse}}\\ 
 \hbox{\tiny{graining}}}}$. In the ultra-strong coupling regime the squeezing 
 property is eliminated altogether since  the
phase coherence properties of the $Q$-distribution, which is necessary for the squeezing to appear, is destroyed in the domain $\lambda \sim \, \omega$ where the interaction modes 
$\{O(x^{n} \widetilde{\Delta})|n = 0, 1, \ldots\}$ of all orders with incommensurate frequencies arise.
We note that the squeezed state described in Figs. \ref{squeezed_state} (c), \ref{polar-plot} is a \textit{nearly pure state} of the oscillator since, at the particular time, its von Neumann entropy  $\left.S^{(+)}\right|_{\omega t=408.5} = 0.0374$ measuring the mixedness  is much lower than the maximal value. This property may also be studied by observing the qubit reduced density matrix. During its time evolution whenever the oscillator reaches a \textit{nearly} pure state, the qubit also evolves  \textit{close} to a pure state of its own. For inspection, we construct the corresponding qubit density matrix via (\ref{density_matrix}-\ref{density_offdiagonal}):
 \beq
  \rho_{\mathcal{Q}}^{(+)}(\omega t=408.5)=
\begin{pmatrix}
 0.6403 & 0.4703+ i \; 0.0552  \\
0.4703 - i \; 0.0552  & 0.3597
\end{pmatrix}.
\label{q_squeeze_den}
 \eeq
Its eigenvalues and the corresponding eigenvectors read: $ \lambda_1=0.9939, \lambda_2=0.0061,$ 
and $| \lambda_1 \rangle =(0.7958+i \; 0.0935 ) | 1 \rangle + 0.5983 |-1\rangle,  
 | \lambda_2 \rangle = (0.5942+ i \; 0.0698 ) | 1 \rangle + 0.8013 |-1\rangle$, respectively. We observe that the qubit overwhelmingly ($99\%$) stays at the pure state $| \lambda_1 \rangle$. Physically, emergence of a squeezed state requires proper phase correlation between the quantum modes. Occurrence of a random statistical mixture comprising of a large number of pure quantum states destroys the phase correlation. Therefore, the emergence of squeezed states during the time evolution of the composite system is possible when the oscillator (qubit) stays \textit{close} to a pure state. 

\par

The Mandel  parameter [\cite{M1979}] is often used to determine the classical-quantum boundary in experiments where photons are directly detected. It captures any sub-Poissonian behavior in the photon statistics, and acts as a witness to nonclassicality:
\beq
\mathcal{Q}_{\mathcal{M}}^{(\pm)}   \equiv \langle \Delta\hat{n}^{2}\rangle^{(\pm)}/ \langle \hat{n}\rangle^{(\pm)} -1.  
\label{mandel}
\eeq
The coherent state with $\mathcal{Q}_{\mathcal{M}} = 0$ characterizes the Poissonian statistics and serves as the benchmark for the classical behavior.
In the current model the interference between the states $|\pm \alpha\rangle$ for small values of the phase space separation 
$\alpha$ produces sub-Poissonian photon statistics $\mathcal{Q}_{\mathcal{M}} < 0$ at certain times during part of the oscillatory cycle. Parallel to the emergence of the squeezed states discussed earlier the sub-Poissonian property is  aided by an initial increase in the coupling since then only the first few nonlinear interaction terms become prominent. In the ultra-strong coupling domain, however, it is diminished  due to the loss of coherence as  nonlinear interaction terms of all orders become active.
\begin{figure}[H]
 \begin{center}
  \includegraphics[width=7cm,height=4cm]{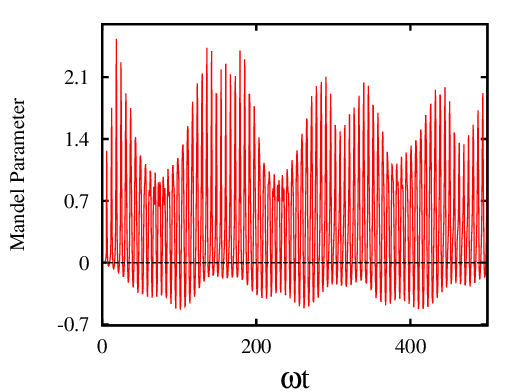}
\caption{The time evolution of the Mandel parameter
for values $\alpha=0.05$, $\lambda=0.6\, \omega, \Delta= 0.15 \,\omega$ and $\epsilon=0.03 \,\omega$.}
\label{mandel_plot}
 \end{center}
\vspace{-0.4cm}
\end{figure}
Moreover, the ultra-strong coupling enhances the expectation value 
of the number of photonic excitations  
(\ref{photon_n1}). The sub-Poissonian statistics is realized when the number of  excitations  
(\ref{photon_n1}) is \textit{not macroscopically large}:  $\langle \hat{n}\rangle^{(\pm)} \lesssim 1$. Using the definition 
(\ref{mandel}) and the expectation values (\ref{photon_n1}, \ref{photon_n2}), we now obtain
$\mathcal{Q}_{\mathcal{M}}^{(\pm)} = \mathcal{O}^{(\pm)}/\langle \hat{n}\rangle^{(\pm)}$,
where the numerator reads
\bea 
\mathcal{O}^{(\pm)}
&=& \frac{2\lambda^2}{\omega^2} |\alpha_{+}|^2
- \frac{2 \lambda}{\omega}\exp(-|\alpha_{+}|^2) \sum_{n=0}^{\infty} 
\frac{|\alpha_{+}|^{2n}}{n!} (n-|\alpha_{+}|^{2})  \mathrm{Re}\lgroup G_{n;1}^{\pm}(t)\rgroup \nn \\
&+&\, \, \frac{\lambda^2}{\omega^2} 
 \exp(-|\alpha_{+}|^2) \left (\sum_{n=0}^{\infty} \frac{|\alpha_{+}|^{2n}}{n!}
\mathrm{Re}\lgroup G_{n;2}^{\pm}\rgroup -\exp(-|\alpha_{+}|^2) \Big [\sum_{n=0}^{\infty} \frac{|\alpha_{+}|^{2n}}{n!}
\mathrm{Re}\lgroup G_{n;1}^{\pm}\rgroup \Big]^2\right). 
\eea
In Fig. \ref{mandel_plot} we find that at various time segments during the oscillatory cycle the Mandel parameter achieves significantly negative values.
Due to the coarse-graining process discussed earlier the quantum interference terms in the $Q$-function is again suppressed and the negativity of the Mandel parameter disappears. However, the negativity of the Mandel parameter is more robust than the squeezing property against chaotic behavior in the phase space in the ultra-strong coupling domain. Another useful characterization of the nonclassicality of the oscillator density matrix is provided by the Wigner $W$-distribution [\cite{Schleich2001}] that may be determined [\cite{CK1993}] via a summation procedure involving the matrix elements of the oscillator density matrix (\ref{o_density_matrix}) in  the displaced number states. 
A more complete  discussion of the observed negativity in the time evolution of the Wigner $W$-distribution in the present model will be given elsewhere.

\section{Conclusion}
\label{Conclusion}
Using the adiabatic approximation we investigate a qubit-oscillator bipartite system in the presence of a static bias term in the Hamiltonian for the strong  and the ultra-strong coupling regimes. Starting with an entangled quasi-Bell state we obtain the evolution of the reduced density matrices of the qubit and the oscillator.  Identifying the series with the Jacobi theta functions in the strong coupling domain we evaluate the qubit density matrix elements in closed form as linear combinations of the theta functions. 
The phase space $Q$-distribution is obtained via the oscillator reduced density matrix. The  
$Q$-function, in turn, provides  
closed form expressions of, say, the first two moments of the quadrature variable as linear combinations of the theta functions. The said evaluations  of various 
physical quantities for the strong coupling  and weak bias limit are  found to be close approximations of their series counterparts which are exact results in the adiabatic scheme. Employing the theta functions, we, in the relevant domain, estimate the revival time of the qubit density matrix and the time of bifurcation of the peaks of the $Q$-distribution observed in Fig. \ref{qfunc_fig}. The differential structure obeyed by the theta functions may also be useful in finding interrelations between various correlation functions appearing  in the study of nonclassicality in photon statistics. Moreover, our approach suggests that a suitable generalization of the theta functions may play an important role for the study of the chaotic state in the ultra-strong coupling domain. 

\par

For an ultra-high qubit-oscillator coupling strength we study the Wehrl entropy and the complexity ${\cal W}_{2}(Q)$ that measure the delocalization of the oscillator in the phase space. A series expression of the complexity ${\cal W}_{2}(Q)$ is obtained. For the coupling range 
$\lambda \lessapprox 0.1\, \omega$ the long-term periodicity is observed to follow from the $O(x^2)$ term in the interaction Hamiltonian. In this regime the $Q$-distribution corresponding to the initial mixed state density matrix of the oscillator evolves at rational fractions of $T_{\mathrm{long}}$ to uniformly separated Gaussian peaks representing `kitten' states. Existence of the interaction-determined multiple time scales in our example, however, gives rise to further bifurcation of Gaussian peaks at times coinciding with the collapse of the off-diagonal qubit density matrix element $\mathrm{Re}(\xi^{(+)})$. In the chaotic $\lambda  \sim \, \omega$ domain the transient production time for the Wehrl entropy has been estimated. Following this period of rapid build up of the Wehrl entropy, the existence of a large number of incommensurate 
interaction-dependent modes leads to a randomization of the phases and a quasi-stationary behavior sets in. For  small values of the phase space amplitude of the quasi-Bell state we observe aspects of the nonclassicality such as existence of  \textit{almost} pure squeezed states of the oscillator, and the negativity of the Mandel parameter. 

\section*{Acknowledgement}
We thank R. Chandrashekar for  discussions and also for pointing out some references to us. One of us (B.V.J.)  acknowledges the support from UGC (India) under the Maulana Azad National Fellowship  scheme. We thank a referee  for making useful suggestions.

\end{document}